%% file: main.tex
\documentclass[conference]{IEEEtran}
\usepackage{tikz}
\usepackage{amsmath}
\usepackage{color}
\usepackage{booktabs}
\usepackage{paralist}
\usepackage{multirow}
\usepackage{longtable}
\usepackage{graphicx}
\usepackage{soul}
\usepackage{footnote}
\usepackage{diagbox}
\usepackage[ruled,linesnumbered]{algorithm2e}

\usepackage{etoolbox}
\usepackage{subfig}

\usepackage{hyperref}

\usepackage{filecontents}
\ifCLASSINFOpdf
\else
\fi
\begin{document}
%
\title{TrojanDam: Detection-Free Backdoor Defense in Federated Learning through Proactive Model Robustification utilizing OOD Data}

\author{\IEEEauthorblockN{Yanbo Dai\IEEEauthorrefmark{2},
            Songze Li\IEEEauthorrefmark{2},
            Zihan Gan\IEEEauthorrefmark{3},
            Xueluan Gong\IEEEauthorrefmark{4}}
      \IEEEauthorblockA{\IEEEauthorrefmark{2}Southeast University}
      \IEEEauthorblockA{\IEEEauthorrefmark{3}HKUST(GZ)}
      \IEEEauthorblockA{\IEEEauthorrefmark{4}Nanyang Technological University}
}
\maketitle

\begin{abstract}
Federated learning (FL) systems allow decentralized data-owning clients to
jointly train a global model through uploading their locally trained updates to
a centralized server. The property of decentralization enables adversaries
to craft carefully designed backdoor updates to make the global model
misclassify only when encountering adversary-chosen triggers. Existing defense
mechanisms mainly rely on post-training detection after receiving updates. These methods
either fail to identify updates which are deliberately fabricated statistically
close to benign ones, or show inconsistent performance in different FL training
stages. The effect of unfiltered backdoor updates will accumulate in the global
model, and eventually become functional. Given the difficulty of ruling out every
backdoor update, we propose a backdoor
defense paradigm, which focuses on proactive robustification on the
global model against potential
backdoor attacks. We first reveal that the successful launching of backdoor
attacks in FL stems from the lack of conflict between malicious and benign updates on redundant neurons of ML models. We proceed to prove the
feasibility of activating redundant neurons utilizing out-of-distribution
(OOD) samples in centralized settings, and migrating to FL settings to propose a
novel backdoor defense mechanism, TrojanDam. The proposed mechanism has the FL
server continuously inject fresh OOD mappings into the global model to activate
redundant neurons, canceling the effect of backdoor updates during aggregation. We conduct
systematic and extensive experiments to illustrate the
superior performance of TrojanDam, over several SOTA backdoor defense methods
across a wide range of FL settings. 
\end{abstract}


%
\IEEEpeerreviewmaketitle

\input{intro}

\input{background}

\input{motivation_method.tex}

\input{exp}

\input{discussion}

\input{conclusion}

\section*{Ethics considerations}
This paper presents work whose goal is to advance the field of backdoor defense
in Federated Learning. Successful deployment of the proposed method could help
to improve the security and robustness of FL systems in various real-world
scenarios, such as in medical image analysis where the integrity of diagnostic
models is paramount, or in industrial control systems where compromised models
could lead to safety hazards. 

Neither the process nor the contributions of this research poses any negative
societal impacts, including but not limited to breaking the security of computer
systems, collecting private data, and violating human rights.

\bibliographystyle{plain}
\bibliography{reference}

\appendix
\section*{Supplementary Experiments}
\subsection{Performance of FreqFed against DBA Attacks}
\label{appendix_dba}
\begin{figure}[!h]
\begin{center}
\centerline{\includegraphics[width=\columnwidth]{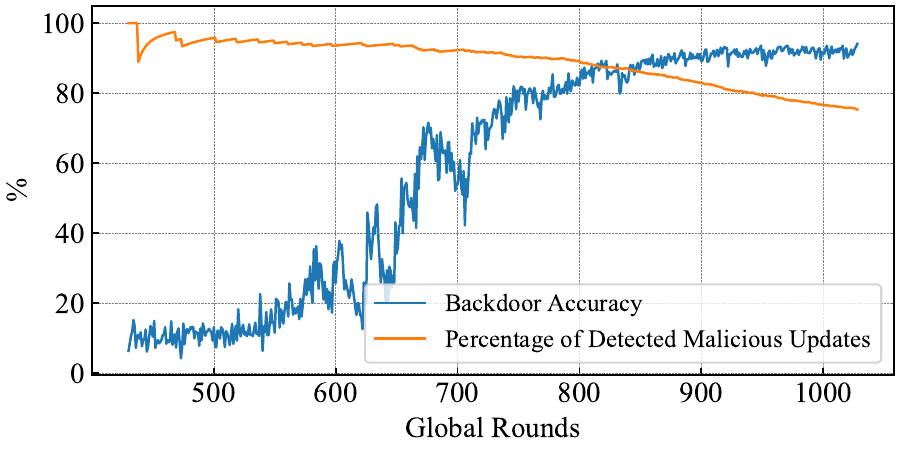}}
\caption{Backdoor accuracy and the percentage of detected malicious updates of
FreqFed against DBA attacks.}
\label{fig:freqfed_dba}
\end{center}
\end{figure}

Figure~\ref{fig:freqfed_dba} illustrates the defense performance of FreqFed
against DBA attacks. Although FreqFed initially detects over 90\% of the
poisoned updates, the backdoor accuracy (BA) still rises to around 40\% by the
600th global round. As training progresses, the detection rate drops below 80\%,
leading to a BA exceeding 90\%. This degradation is due to the strong influence
of backdoor updates trained with DBA: even a small fraction of undetected
malicious updates can quickly embed backdoors into the global model. As more
poisoned updates bypass the defense, detection becomes increasingly difficult,
exacerbating the attack’s effectiveness over time.

\subsection{Complete Empirical Results}
\label{appendix}
We further provide complete experiment results of the backdoor defense
performance of all evaluated methods on CIFAR100, EMNIST, ResNet34 and VGG16 in
Table \ref{tab:resnet34}, \ref{tab:vgg16}, \ref{tab:cifar100_appendix},
\ref{tab:cifar100_alphas}, \ref{tab:emnist_appendix}, \ref{tab:emnist_alpha}.
All presented empirical results illustrate the effectiveness of TrojanDam,
which successfully limits the BA to comparable performance with random guesses and
achieves among the best performance in comparison with the evaluated SOTA backdoor
defense mechanisms.

\begin{table*}[htbp]
  \centering
  \caption{BA (MA) of all evaluated methods against
  single client attack with different combinations of backdoor type and
  malicious training algorithm. \textbf{Bold} values indicate the lowest
  metrics, and \underline{underlined} values represents the second lowest
  ones. The evaluated model architecture is ResNet34. The poisoning lasts for
  400 global rounds.}
  \scalebox{0.77}{
    \begin{tabular}{cc||cccccccccc}
    \toprule
    training alg. & bkdr types & Nodefense & Deepsight & Foolsgold & FLAME &
    FreqFed & BayBFed & MESAS & FLTrust & Indicator & TrojanDam \\
    \midrule
    \multirow{4}[2]{*}{PGD} & blended & 76.41 (90.03) & \underline{39.24} (90.93) & 72.76 (89.45) & 86.68 (89.23) & 62.31 (90.40) & 71.22 (89.84) & 60.24 (90.32) & 61.24 (88.22) & 55.44 (90.30) & \textbf{28.2} (89.46) \\
          & semantic & 60.27 (90.09) & 60.70 (90.74) & 62.97 (90.53) & 66.54 (89.45) & \underline{43.17} (90.83) & 79.39 (89.53) & 90.24 (90.12) & 57.34 (89.51) & 55.04 (90.31) & \textbf{0.00} (88.69) \\
          & edge case & 46.02 (90.11) & 34.10 (90.14) & 46.28 (90.73) & 68.09 (89.67) & 42.61 (89.23) & 51.24 (90.29) & 44.30 (89.89) & 33.45 (89.69) & \underline{19.82} (90.53) & \textbf{11.59} (89.30) \\
          & TacT  & 48.28 (90.03) & 48.10 (90.88) & 86.45 (90.46) & 39.33 (89.20) & 93.81 (90.98) & 81.09 (89.33) & 78.19 (90.19) & 92.67 (89.50) & \underline{39.10} (89.73) & \textbf{8.02} (88.74) \\
    \midrule
    \multirow{4}[2]{*}{Neurotoxin} & blended & 77.00 (90.97) & 38.79 (90.84) & 75.26 (90.62) & 88.95 (89.35) & \textbf{19.29} (88.40) & 63.04 (89.61) & 64.12 (90.35) & 61.23 (88.44) & 53.91 (90.38) & \underline{27.72} (88.34) \\
          & semantic & 68.18 (90.93) & 61.38 (90.76) & 70.95 (90.59) & 69.58 (89.40) & \textbf{0.00} (90.12) & 78.08 (89.74) & 87.05 (90.15) & 59.96 (88.54) & 26.85 (90.22) & \underline{0.19} (89.22) \\
          & edge case & 50.08 (90.97) & 36.34 (90.30) & 56.88 (88.69) & 70.00 (89.69) & \underline{14.70} (90.47) & 41.85 (89.39) & 36.44 (89.19) & 32.40 (89.81) & 31.81 (90.51) & \textbf{9.71} (89.02) \\
          & TacT  & 60.60 (90.94) & 50.23 (90.06) & 92.77 (90.33) & \underline{0.36} (90.04) & 87.60 (89.88) & 66.50 (89.38) & 59.41 (90.07) & 89.55 (88.83) & \textbf{0.34} (90.29) & 18.35 (88.91) \\
    \midrule
    \multirow{4}[2]{*}{Chameleon} & blended & 61.56 (90.02) & 35.88 (90.96) & 54.98 (89.57) & 78.12 (89.39) & \textbf{24.19} (90.71) & 50.10 (89.89) & 59.01 (90.26) & 50.45 (89.55) & 45.80 (90.36) & \underline{29.91} (88.39) \\
          & semantic & 61.67 (90.95) & \underline{32.88} (90.78) & 61.20 (90.53) & 67.07 (89.39) & 54.63 (90.53) & 38.73 (89.43) & 74.01 (90.02) & 41.94 (89.51) & 46.04 (90.29) & \textbf{0.00} (88.39) \\
          & edge case & 26.77 (90.04) & \underline{18.04} (91.03) & 28.92 (90.36) & 46.94 (89.31) & 18.10 (90.06) & 25.40 (89.46) & 28.37 (89.89) & 21.07 (90.30) & 26.84 (90.34) & \textbf{13.70} (89.18) \\
          & TacT  & 78.62 (89.23) & 72.13 (90.85) & 77.63 (90.28) & 65.20 (90.25) & \underline{28.98} (90.07) & 66.09 (89.59) & 85.36 (90.04) & 57.35 (89.90) & 30.55 (90.47) & \textbf{5.63} (88.87) \\
    \bottomrule
    \end{tabular}%
  }
  \label{tab:resnet34}%
\end{table*}%

\begin{table*}[htbp]
  \centering
  \caption{BA (MA) of all evaluated methods against
  single client attack with different combinations of backdoor type and
  malicious training algorithm. \textbf{Bold} values indicate the lowest
  metrics, and \underline{underlined} values represents the second lowest
  ones. The evaluated model architecture is VGG16. The poisoning lasts for
  300 global rounds.}
  \scalebox{0.75}{
    \begin{tabular}{cc||cccccccccc}
    \toprule
    training alg. & bkdr types & Nodefense & Deepsight & Foolsgold & FLAME &
    FreqFed & BayBFed & MESAS & FLTrust & Indicator & TrojanDam \\
    \midrule
    \multirow{3}[2]{*}{PGD} & blended & 83.47 (90.17) & 43.22 (90.94) &
    81.20 (90.07) & 90.23 (89.57) & 60.84 (90.52) & 71.10 (89.20) & 68.11 (90.66) & 59.46 (90.48) & \underline{20.20} (90.95) &
    \textbf{13.21} (89.05) \\
          & semantic & 85.35 (90.08) & 86.45 (90.22) & 86.37 (89.85) &
          \underline{70.40} (89.57) & 54.61 (90.13) & 88.71 (88.66) & 76.90 (90.47) & 67.74 (89.38) & 76.22 (90.76) &
          \textbf{18.64} (89.89) \\
          & edge case & 70.95 (90.38) & 54.61 (90.29) & 74.57 (90.00) &
          89.49 (89.72) & 56.13 (90.35) & 62.80 (89.18) & 64.88 (90.22) & 45.59 (90.75) & \textbf{5.96} (90.93) & \underline{25.13} (90.34)
          \\
    \midrule
    \multirow{3}[2]{*}{Neurotoxin} & blended & 85.12 (90.18) & 44.58 (90.05) &
    82.68 (90.12) & 94.17 (89.29) & 60.42 (90.56) & 65.97 (89.20) & 62.13 (90.75) & 58.68 (89.25) & \underline{35.51} (90.73) &
    \textbf{25.40} (89.08) \\
          & semantic & 92.81 (90.11) & 85.56 (90.17) & 89.26 (89.92) &
          76.52 (88.47) & \underline{53.63} (89.96) & 73.16 (88.55) & 94.51 (89.16) & 75.41 (90.41) & 73.39 (90.65) &
          \textbf{0.48} (88.70) \\
          & edge case & 78.91 (90.22) & 59.57 (90.13) & 78.81 (89.95) &
          88.61 (89.63) & 46.79 (90.35) & 64.75 (88.63) & 59.89 (90.11) & 52.23 (90.51) & \underline{11.61} (89.82) & \textbf{8.19} (89.15)
          \\
    \midrule
    \multirow{3}[2]{*}{Chameleon} & blended & 89.83 (90.28) & 49.69 (90.19) &
    87.92 (90.06) & \textbf{10.13} (90.35) & \underline{17.85} (90.56) & 76.88 (89.29) & 78.78 (90.26) & 69.26 (90.55) & 39.93 (90.56) &
    28.76 (88.76) \\
          & semantic & 79.53 (90.19) & 70.57 (90.13) & 71.15 (90.14) &
          \textbf{0.00} (90.37) & 27.72 (90.65) & 77.47 (89.28) & 86.60 (89.09) & 71.60 (89.93) & 78.89 (89.83) & \underline{7.21} (90.26)
          \\
          & edge case & 81.02 (90.18) & 67.85 (90.10) & 80.83 (89.89) &
          \textbf{2.84} (89.20) & 8.16 (90.91) & 53.32 (89.48) & 64.42 (90.44) & 54.46 (90.13) & 20.62 (90.82) & \underline{7.34} (89.69)
          \\
    \bottomrule
    \end{tabular}%
  }
  \label{tab:vgg16}%
\end{table*}%

\begin{table*}[htbp]
  \centering
  \caption{BA (MA) of all evaluated methods against
  single client attack with different combinations of backdoor type and
  malicious training algorithm. \textbf{Bold} values indicate the lowest
  metrics, and \underline{underlined} values represents the second lowest
  ones. The evaluated dataset is CIFAR100. The poisoning lasts for
  400 global rounds.}
  \scalebox{0.75}{
    \begin{tabular}{cc||cccccccccc}
    \toprule
    training alg. & bkdr types & Nodefense & Deepsight & Foolsgold & FLAME &
    FreqFed & BayBFed & MESAS & FLTrust & Indicator & TrojanDam \\
    \midrule
    \multirow{3}[2]{*}{PGD} & blended & 66.72 (68.53) & 28.72 (68.13) &
    63.35 (68.12) & 81.97 (66.18) & 62.73 (66.57) & 69.40 (66.57) & 66.07 (66.19) & 45.93 (65.65) & \underline{13.07} (66.35) &
    \textbf{0.22} (64.78) \\
          & edge case & 81.84 (68.60) & 66.85 (67.88) & 81.29 (68.02) &
          92.07 (66.27) & 86.95 (65.24) & 84.47 (66.65) & 79.37 (66.37) & 76.80 (66.03) & \textbf{0.00} (67.38) & \underline{1.14} (64.45)
          \\
          & TacT  & 97.51 (67.84) & \underline{80.41} (67.70) & 97.39 (67.71)
          & 100.00 (65.18) & 78.63 (65.02) & 96.65 (66.54) & 95.82 (66.75) & 85.81 (66.80) & \textbf{0.00} (67.23) & \textbf{0.00} (65.21)
          \\
    \midrule
    \multirow{3}[2]{*}{Neurotoxin} & blended & 65.20 (68.55) &
    \underline{27.76} (68.41) & 63.78 (68.30) & 63.78 (66.24) &
    60.92 (65.14) & 66.44 (66.63) & 66.39 (66.38) & 56.61 (67.03) & 52.31 (67.28) & \textbf{0.31} (67.55) \\
          & edge case & 83.83 (68.40) & 69.33 (68.05) & 80.99 (67.95) &
          92.09 (66.39) & 85.71 (65.57) & 85.21 (66.66) & 78.66 (66.72) & 77.83 (67.45) & \underline{21.38} (67.47) & \textbf{1.79} (65.72)
          \\
          & TacT  & 96.86 (67.72) & \underline{80.44} (67.69) & 97.02 (67.69)
          & \textbf{0.00} (66.41) & 80.32 (64.59) & 96.68 (66.38) & 91.08 (66.09) & 81.90 (65.24) & \textbf{0.00} (67.13) &
          \textbf{0.00} (63.68) \\
    \midrule
    \multirow{3}[2]{*}{Chameleon} & blended & 28.21 (68.24) & 16.49 (68.30) &
    26.24 (68.29) & 54.68 (66.30) & 19.89 (66.37) & 23.46 (66.30) & 24.28 (66.89) & 17.25 (65.33) & \underline{12.12} (67.21) &
    \textbf{0.01} (66.35) \\
          & edge case & 42.28 (68.52) & 33.85 (68.33) & 39.74 (68.20) &
          62.08 (66.28) & 42.78 (64.96) & 47.05 (66.22) & 35.62 (66.74) & 27.88 (64.03) & \textbf{0.17} (67.12) & \underline{3.70} (67.33)
          \\
          & TacT  & 90.10 (67.74) & \underline{48.86} (67.78) & 86.76 (67.66)
          & 98.68 (65.08) & 45.01 (64.12) & 54.52 (66.61) & 60.86 (66.50) & 60.17 (66.66) & \textbf{0.00} (67.21) & \textbf{0.00} (64.21)
          \\
    \bottomrule
    \end{tabular}%
    }
  \label{tab:cifar100_appendix}%
\end{table*}%

\begin{table*}[htbp]
  \centering
  \caption{BA (MA) of all evaluated methods against single
  client attack under different non-IID settings, and different poisoned
  learning rates (\textit{plr}s) adopted by adversaries. The considered backdoor type,
  and malicious training algorithm are blended backdoors and Neurotoxin
  respectively. The evaluated dataset is CIFAR100. \textbf{Bold} values indicate
  the lowest metrics, and \underline{underlined} values represents the second
  lowest ones.}
  \scalebox{0.78}{
    \begin{tabular}{cc||cccccccccc}
    \toprule
    alpha & poisoned lr & Nodefense & Deepsight & Foolsgold & FLAME & FreqFed & BayBFed & MESAS & FLTrust& Indicator & TrojanDam \\
    \midrule
    \multirow{3}[2]{*}{0.9} & 0.01  & 53.43 (67.58) & 23.93 (67.91) &
    52.04 (67.71) & 68.87 (65.12) & 50.35 (65.76) & 53.00 (66.23) & 50.45 (66.76) & 40.32 (66.32) & \underline{13.51} (67.34) &
    \textbf{0.29} (67.21) \\
          & 0.025 & 65.20 (68.55) & \underline{27.76} (68.41) & 63.78 (68.30)
          & 63.78 (66.24) & 60.92 (65.14) & 66.44 (66.63) & 66.39 (66.38) & 56.61 (67.03) & 52.31 (67.28) & \textbf{0.31} (67.55) \\
          & 0.04  & 73.10 (68.56) & 28.90 (67.89) & 70.85 (68.29) &
          \textbf{0.16} (65.92) & 5.38 (65.99) & 74.14 (66.73) & 76.37 (66.53) & 67.04 (65.82) & 63.01 (67.31) & \underline{0.70} (65.03)
          \\
    \midrule
    \multirow{3}[2]{*}{0.5} & 0.01 & 53.46 (68.10) & 28.45 (67.83) &
    50.79 (67.87) & 67.97 (65.34) & 46.34 (65.43) & 56.36 (66.01) & 51.42 (66.83) & 41.14 (66.44) & \underline{0.34} (67.62) &
    \textbf{0.07} (66.48) \\
          & 0.025 & 67.85 (68.00) & 31.58 (68.00) & 64.85 (67.87) &
          78.45 (65.15) & 41.88 (65.30) & 68.07 (65.55) & 66.89 (65.92) & 55.47 (65.99) & \underline{5.67} (67.42) & \textbf{0.66} (66.73)
          \\
          & 0.04  & 64.59 (68.05) & 27.02 (67.78) & 63.31 (68.06) &
          \underline{0.23} (66.34) & 39.68 (65.67) & 69.94 (66.35) & 68.47 (66.89) & 64.76 (65.92) & 18.76 (67.31) & \textbf{0.18} (67.07)
          \\
    \midrule
    \multirow{3}[2]{*}{0.2} & 0.01  & 59.90 (66.75) & 28.27 (66.56) &
    56.72 (66.76) & 71.81 (62.53) & 59.49 (60.73) & 47.45 (64.17) & 61.25 (64.51) & 42.37 (62.00) & \underline{0.24} (66.14) &
    \textbf{0.11} (65.05) \\
          & 0.025 & 74.01 (66.92) & 32.71 (66.69) & 56.24 (66.77) &
          82.58 (62.50) & 70.03 (59.66) & 65.98 (64.40) & 69.74 (64.93) & 59.66 (64.50) & \underline{10.01} (66.06) & \textbf{0.12} (64.47)
          \\
          & 0.04  & 81.24 (66.80) & 34.01 (66.86) & 79.60 (66.71) &
          \textbf{0.25} (63.98) & 19.13 (60.22) & 79.23 (64.91) & 79.88 (64.86) & 70.64 (64.04) & 14.78 (66.07) & \underline{1.91} (64.52)
          \\
    \bottomrule
    \end{tabular}%
  }
  \label{tab:cifar100_alphas}%
\end{table*}%

\begin{table*}[htbp]
  \centering
  \caption{BA (MA) of all evaluated methods against
  single client attack with different combinations of backdoor type and
  malicious training algorithm. \textbf{Bold} values indicate the lowest
  metrics, and \underline{underlined} values represents the second lowest
  ones. The evaluated dataset is EMNIST. The poisoning lasts for
  300 global rounds.}
  \scalebox{0.73}{
    \begin{tabular}{cc||cccccccccc}
    \toprule
    training alg. & bkdr types & Nodefense & Deepsight & Foolsgold & FLAME & FreqFed &BayBFed & MESAS & FLTrust & Indicator & TrojanDam \\
    \midrule
    PGD   & pixel-pattern & 100.00 (99.71) & 99.99 (99.69) & 100.0 (99.70) &
    \textbf{10.01} (99.67) & 100.00 (99.61) & 100.00 (99.65) & 100.00 (99.70) & 99.99 (99.67) & 66.23 (99.70) & \underline{10.27} (99.72)
    \\
    Neurotoxin & pixel-pattern & 100.0 (99.71) & 100.00 (99.69) &
    100.00 (99.70) & \textbf{10.01} (99.69) & 100.00 (99.65) & 100.00 (99.67) & 100.00 (99.69) & 99.99 (99.67) & 15.17 (99.70) &
    \underline{12.51} (99.48) \\
    Chameleon & pixel-pattern & 100.00 (99.68) & 100.00 (99.69) &
    100.00 (99.70) & \textbf{10.01} (99.68) & 38.02 (99.64) & 100.00 (99.66) & 100.00 (99.68) & 100.00 (99.65) & 65.78 (99.68) &
    \underline{18.58} (99.50) \\
    \bottomrule
    \end{tabular}%
  }
  \label{tab:emnist_appendix}%
\end{table*}%

\begin{table*}[htbp]
  \centering
  \caption{BA (MA) of all evaluated methods against single
  client attack under different non-IID settings, and different poisoned
  learning rates (\textit{plr}s) adopted by adversaries. The considered backdoor type,
  and malicious training algorithm are blended backdoors and Neurotoxin
  respectively. The evaluated dataset is EMNIST. \textbf{Bold} values indicate
  the lowest metrics, and \underline{underlined} values represents the second
  lowest ones.}
  \scalebox{0.75}{
    \begin{tabular}{cc||cccccccccc}
    \toprule
    alpha & poisoned lr & Nodefense & Deepsight & Foolsgold & FLAME & FreqFed
    & BayBFed & MESAS & FLTrust & Indicator & TrojanDam \\
    \midrule
    \multirow{3}[2]{*}{0.9} & 0.01  & 100.00 (99.71) &
    \underline{99.99} (99.75) & 100.00 (99.71) & \textbf{10.01} (99.65) &
    99.99 (99.59) & 100.00 (99.66) & 99.98 (99.69) & 99.99 (99.64) & \textbf{10.01} (99.72) & \textbf{10.01} (99.45) \\
          & 0.025 & 100.00 (99.71) & 99.99 (99.69) & 100.00 (99.70) &
          \textbf{10.00} (99.69) & 100.00 (99.65) & 100.00 (99.67) & 100.00 (99.69) & 99.99 (99.67) & 13.29 (99.70) &
          \underline{10.33} (99.48) \\
          & 0.04  & 100.00 (99.70) & 100.00 (99.76) & 100.00 (99.71) &
          \textbf{10.00} (99.67) & 100.00 (99.61) & 99.98 (99.64) & 99.99 (99.65) & 100.00 (99.69) & 41.19 (99.71) &
          \underline{10.75} (99.52) \\
    \midrule
    \multirow{3}[2]{*}{0.5} & 0.01  & 99.98 (99.67) & 98.45 (99.71) &
    99.99 (99.69) & 10.67 (99.67) & 99.94 (99.54) & 99.96 (99.65) & 99.95 (99.66) & 99.60 (99.64) & \textbf{10.00} (99.66) &
    \underline{10.11} (99.43) \\
          & 0.025 & 100.00 (99.67) & 99.81 (99.70) & 99.98 (99.67) &
          \textbf{10.01} (99.67) & 99.91 (99.55) & 100.00 (99.65) & 99.99 (99.66) & 99.92 (99.65) & \underline{10.04} (99.66) &
          10.14 (99.35) \\
          & 0.04  & 100.00 (99.66) & 99.30 (99.71) & 100.00 (99.68) &
          \textbf{10.01} (99.68) & 99.99 (99.60) & 99.99 (99.64) & 99.98 (99.57) & 99.96 (99.57) & \underline{10.03} (99.67) &
          10.25 (99.47) \\
    \midrule
    \multirow{3}[2]{*}{0.2} & 0.01  & 99.77 (99.61) & 98.89 (99.68) &
    99.45 (99.62) & 99.98 (99.66) & 98.84 (99.12) & 94.27 (99.64) & 92.05 (99.41) & 96.87 (99.55) & \textbf{10.04} (99.66) &
    \underline{12.40} (99.53) \\
          & 0.025 & 99.90 (99.60) & 99.27 (99.68) & 99.43 (99.62) &
          \textbf{10.02} (99.64) & 99.97 (99.50) & 96.63 (99.64) & 99.90 (99.40) & 99.23 (99.46) & \underline{10.03} (99.66) &
          12.54 (99.42) \\
          & 0.04  & 99.93 (99.58) & 99.30 (99.67) & 99.88 (99.61) &
          \textbf{10.02} (99.63) & 99.97 (99.15) & 99.95 (99.59) & 99.94 (99.60) & 99.65 (99.22) & \underline{10.04} (99.65) &
          12.34 (99.06) \\
    \bottomrule
    \end{tabular}%
  }
  \label{tab:emnist_alpha}%
\end{table*}%

\end{document}

%% file: intro.tex
\section{Introduction}

The proliferation of personal and corporate data brought by the booming of
computational resources makes proper exploitation on sensitive data a
challenging problem. Federated learning (FL) \cite{mcmahan2017communication}
offers a privacy-preserving solution through enabling multiple data owners to
jointly train a global model under the coordination of a central server. For
every global round in FL, the server first broadcasts the global model to
selected local clients. The server proceeds to aggregate received updates, which
are trained by participating clients on their local datasets, to generate a
global model for the next iteration. During the interaction between the server
and clients, only the model updates, instead of their raw data, are exposed.

Despite respecting participants' privacy, FL frameworks are notorious for their
vulnerability against various malicious behaviors \cite{bagdasaryan01,
Zhengming01, fang2020local, geiping2020inverting, fung2020limitations,
nasr2019comprehensive, bhagoji2019analyzing}. Due to the decentralized nature of
the FL, adversaries could easily compromise participants to launch either
untargeted attacks \cite{fang2020local, shejwalkar2021manipulating,
chen2017distributed, guerraoui2018hidden} or backdoor attacks
\cite{bhagoji2019analyzing, bagdasaryan01}. Different from untargeted attackers,
who aim to compromise the overall performance of the global model, backdoor
adversaries are especially destructive because of their stealth. Once
successfully injected, the infected model will misclassify into an
adversary-chosen label when encountering predefined triggers, while leaving
other tasks uninfluenced. The feasibility of injecting backdoors into FL models
has been extensively demonstrated in prior work. Bagdasaryan et \textit{al.}
\cite{bagdasaryan01} proposes to upload a poisoned model, which is scaled up to
cancel the effect of other benign updates during model aggregation, to
successfully replace a backdoor model with the global model. Zheng et
\textit{al.} \cite{Zhengming01} identifies parameters which are frequently
updated by benign updates, and excludes them from training the poisoned model to
inject more stealthy and durable backdoors. 

\begin{figure}[!t]
    \begin{center}
    \centerline{\includegraphics[width=\columnwidth]{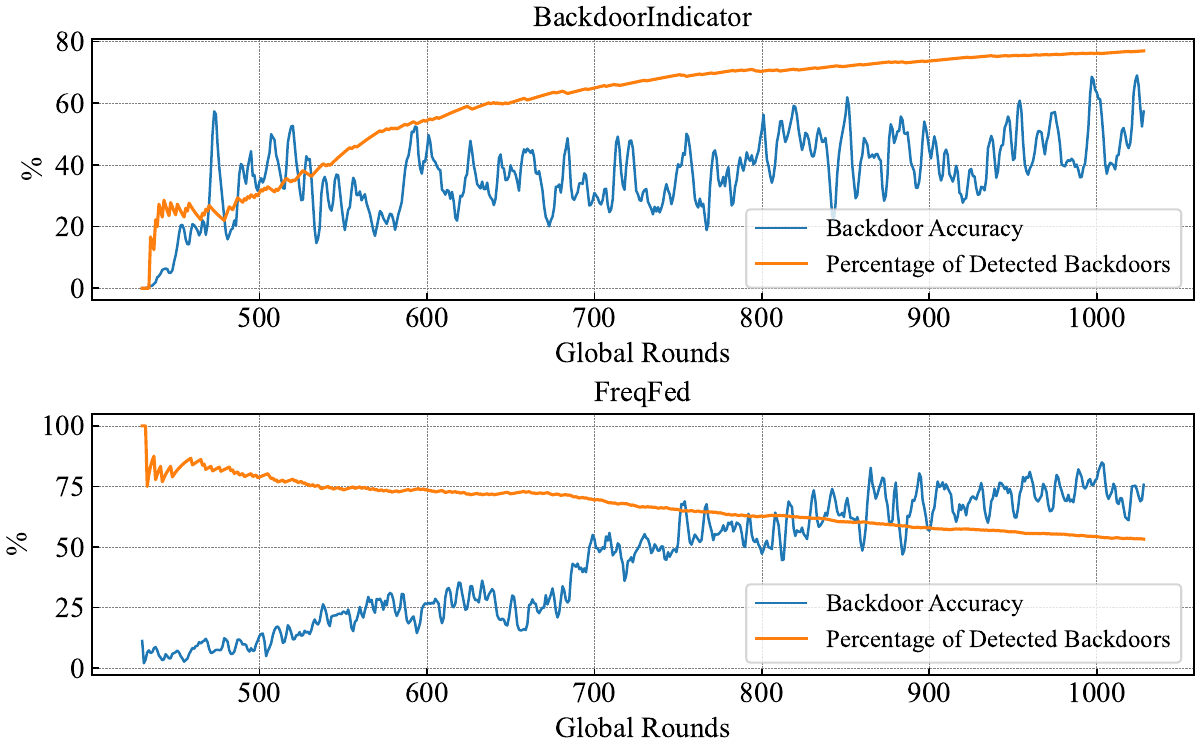}}
    \caption{The backdoor task accuracy and the percentage of detected backdoors
    of \textbf{(UPPER)} BackdoorIndicator, and \textbf{(LOWER)} FreqFed.}
    \label{fig:motivation_against_longterm}
    \end{center}
    \vspace{-0.4in}
\end{figure}

The threat posed by backdoor attacks necessitates designs of backdoor
elimination mechanism in practical FL frameworks. 
Previous works on defending against backdoor attacks in FL rely on
\emph{post-processing} techniques. After collecting updates from clients, the
server either tries to limit the influence of backdoor updates on the global
model \cite{zitengsun01, cao2021provably, McMahan2017LearningDP,
Naseri2020LocalAC, Yin2018ByzantineRobustDL}, or identify and further filter out
backdoors through comparing model parameters\cite{rieger2022deepsight,
blanchard2017machine, nguyen2022flame, foolsgold, wang2022rflbat,
munoz2019byzantine, shen2016auror, zhao2020shielding, kumari2023baybfed,
cao2020fltrust, fereidooni2023freqfed, krauss2023mesas, rieger2022crowdguard}.
However, it is widely believed that influence-reduction-based methods can merely
slow the rate of backdoor success, rather than eliminating the injected
backdoors entirely. For instance, Sun et \textit{al.} \cite{zitengsun01}
proposes to clip all received updates to an agreed bound to limit the influence
of scaled backdoor updates. Other approaches apply differential privacy by
injecting random noise into the aggregated model, aiming to obscure adversarial
features. Detect-then-filter methods generally assume that poisoned updates
differ from benign ones in the parameter space, using statistical metrics to
flag anomalies. Updates exhibiting significant deviations are treated as
poisoned and excluded from aggregation. A recent work, BackdoorIndicator
\cite{backdoorindicator}, enhances post-training detection by proactively
injecting pseudo-backdoors into the global model prior to client distribution.
When adversaries later upload malicious updates, the pre-planted
pseudo-backdoors are triggered, allowing the server to detect and remove
compromised updates. While BackdoorIndicator achieves SOTA detection
performance, it struggles to identify poisoned updates during the early stages
of FL training. Consequently, some malicious updates are aggregated into the
global model, allowing the backdoor to gradually take effect as the adversary
continues participating in training. We empirically evaluate the performance of
BackdoorIndicator and another SOAT backdoor detection scheme, FreqFed
\cite{fereidooni2023freqfed}, against long-term backdoor injection. Assuming the
adversary continuously attacks throughout training, our results (Figure
\ref{fig:motivation_against_longterm}) show that BackdoorIndicator fails to
detect early-stage attacks, allowing backdoor accuracy to rise to ~40\% before
detection stabilizes. Although FreqFed maintains an 80\% detection rate in the
first 100 rounds, unfiltered malicious updates gradually influence the global
model. As the poisoned models become less distinguishable from benign ones,
detection effectiveness declines over time. \emph{These observations highlight
the limitations of post-hoc detection methods and underscore the need for
defense mechanisms that do not rely solely on discriminating between benign and
backdoor updates. Instead, robust defenses must be capable of withstanding
persistent backdoor injection across long FL training horizons.}

\begin{figure}[!t]
    \begin{center}
    \centerline{\includegraphics[width=\columnwidth]{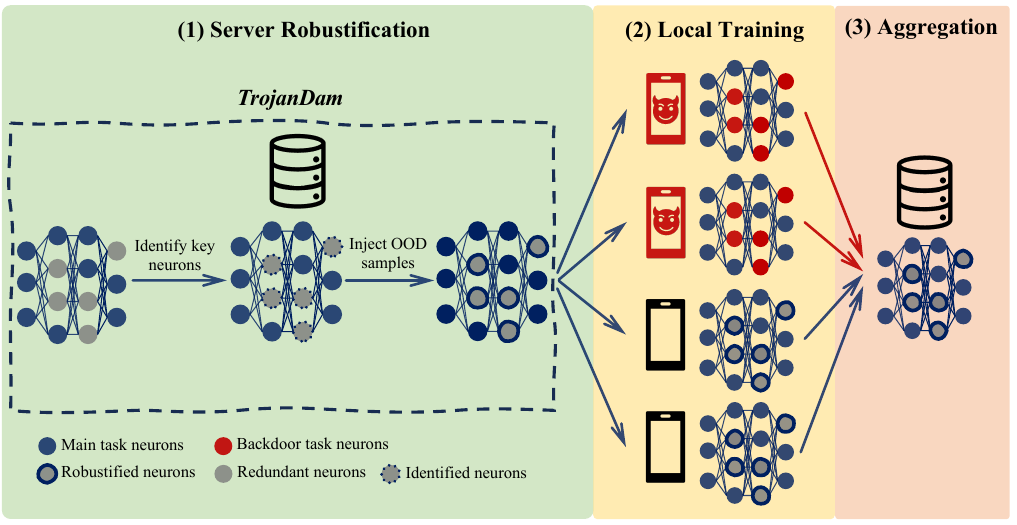}}
    \caption{The overview of FL systems with TrojanDam.}
    \label{fig:overview}
    \end{center}
    \vspace{-0.3in}
\end{figure}

In this work, we propose a novel \emph{pre-processing} backdoor defense mechanism,
\emph{TrojanDam}, to defend against backdoor injection by robustifying the FL
global model \textit{prior} to broadcasting. \textbf{Unlike exitsing approaches,
TrojanDam does not require the server to identify potential malicious updates
from clients.} Instead, we propose to have the server
\emph{proactively robustify redundant neurons}, which are the most susceptible
to backdoor injections. 
The key observation is that effective FL backdoor attacks tend to exploit
redundant neurons which are rarely updated by benign training. Fortunately,
these neurons could be activated to mitigate backdoors, by training using a
mixture of main task samples and OOD samples, which we refer to as flood data.
TrojanDam operates by activating redundant neurons at the beginning of every
global round using flood data, thereby reducing their capacity to encode
adversarial triggers. Since the server typically lacks access to main task data,
we introduce an additional set of OOD samples-termed shadow data-to approximate
the distribution of the main task. The server computes gradients from a mixture
of flood and shadow data, and projects them onto key kernels identified as
influential to redundant neurons. This targeted gradient projection allows for a
more efficient and lightweight update, reducing the required amount of flood
data and improving deployment practicality. We demonstrate the overview of FL
systems equipped with TrojanDam in Figure \ref{fig:overview}. At the beginning
of every FL global round, the server activates redundant neurons by introducing
OOD mappings before broadcasting. While adversaries may still attempt to inject
backdoors by exploiting these neurons, their efforts are mitigated by
aggregation with benign updates in which redundant neurons remain activated

We further conduct extensive experiments on three image datasets: CIFAR10,
CIFAR100 \cite{krizhevsky2009learning} and EMNIST \cite{cohen2017emnist}, with
three model architectures: VGG16 \cite{simonyan2014very}, ResNet18 and ResNet34
\cite{he2016deep}. We demonstrate the effectiveness of TrojanDam, by comparing
its backdoor suppression performance with several SOTA defense mechanisms
against a powerful adversary. The adversary could upload different types of
backdoor updates trained using different training algorithms. We assume that the
adversary could continuously participate in the FL paradigm for a large number
of global rounds. We also provide results to reveal the influence of several key
hyper-parameters, including the source of the flood dataset, flood dataset size,
and the ratio of key kernels, on the performance of TrojanDam.

In summary, our contribution is four folds: 1) we propose a novel
detection-free backdoor defense paradigm, which mainly relies on a
proactive process on the global model to robustify the FL model. 2) We reveal
that redundant neurons in neural networks could be robustified in a centralized
setting through injecting OOD samples. 3) Motivated by the above observation, we
migrate such an idea to FL settings, and propose a novel backdoor suppression
mechanism, TrojanDam. The proposed method has the FL server consistently inject
OOD mappings to robustify redundant neurons in the global model, canceling
backdoors during aggregation. 4) We provide extensive empirical results to
demonstrate the effectiveness of TrojanDam over several SOTA backdoor defense
methods across various adversarial and FL scenarios.

%% file: background.tex
\section{Preliminaries and Related Work}
\subsection{Federated Learning}
In an FL system, multiple data-owning clients jointly train
a global DNN model under the guidance of a central server. Participating clients
interact with the central server through downloading the global model, and
uploading local models trained on their private data. This training paradigm
reduces the infrastructure cost through offloading computing tasks to local
devices, and also preserves users' privacy as no raw data is exchanged
throughout the whole training process. The baseline algorithm for implementing an FL system is FedAVG \cite{mcmahan2017communication}. Generally speaking,
FedAVG aims to minimize the summation of the local empirical losses
$\sum_{i=1}^S\mathcal{L}_i(\theta)$ of $S$ participating clients. Here, we
denote $\mathcal{L}_i$ as the cross-entropy over the local dataset $D_i$ of
client $i$, and $\theta$ as the global model. For global round $t$ of the FL,
the server first broadcasts the current FL global model $\theta^t$ to a subset
$S_t$ of selected clients. Each local client $i$ in $S_t$ then initializes its local
model $\theta_i^t$ from $\theta^t$, and then trains its local model through
computing updates on $D_i$. The server then aggregates all received local
updates through $\theta^{t+1}=\frac{1}{|S^t|}\sum_{i\in S_t}\theta_i^t$ to
generate the global model for global round $t+1$. In the rest of the paper, we
adopt FedAVG for FL training.

\subsection{Backdoor Attacks in FL}
After corrupting local clients, the adversary could incorporate
poisoned samples into the local training dataset.
The adversary proceeds to train the poisoned model through computing updates on
the constructed training dataset using mini-batch stochastic gradient descent.
Besides the baseline backdoor injection methods, previous works have proposed
various advanced backdoor training methods, which enjoy stronger stealth against
backdoor defenses. To escape from the norm-clipping defense, which limits the
influence of individual updates through regularizing the norm of each received
update to a predefined bound, the adversary could adopt projected gradient
descend (PGD) to inject the backdoor \cite{zitengsun01}. Specifically, the
poisoned model is trained and then projected onto an $\ell_2$ ball around the
model of the previous iteration. 

Another line of work tries to fabricate backdoor updates to make them
statistically close to benign updates to escape from backdoor detection. F3BA
\cite{fang2023vulnerability} first identifies a small fraction of parameters
with the lowest movement-based importance scores computed from the element-wise
product between their weights and gradients. The identified candidate parameters
are the least important to the performance of the main task. The adversary then
compromises these parameters by flipping their sign to enhance their sensitivity
to the trigger. AutoAdapt proposes to leverage the Augmented Lagrangian-based
methods for automatically evading backdoor detection \cite{krauss2024automatic}.
The adversary could first train several benign models to compute legitimate
values for the detection metric used by the known defense. The extracted values
further form a valid range, which is then transformed into the Lagrangian
multipliers. The backdoor model is then trained under the constraint of these
Lagrangian multipliers to achieve a satisfactory solution. Several recent
works propose to cast the backdoor attack as a joint optimization problem. While
training poisoned local models, they directly optimize the backdoor trigger to
make models misclassify into the adversarial target label. CerP \cite{lyu2023poisoning} formulates the distributed backdoor attack
in terms of three learning objectives, which are the fine-tuning of backdoor
triggers, the control over poisoned model bias, and the diversity of poisoned
local models. PFedBA \cite{lyu2024lurking} further improves the trigger
optimization procedure through aligning gradients of the backdoor task with
those of the benign task.

Adversaries could also choose to inject different types of backdoors into the
global model. BadNets \cite{gu2020badnets} proposes to directly modify pixels in
the original image, and considers the overlaid pixel-pattern as the backdoor
trigger. Targeted contamination attack (TaCT) \cite{tang2021demon} further
strengthens the stealth of the injected pixel-pattern backdoor by only
assigning the target label to trigger-carrying samples from the specific class.
Blended backdoors \cite{chen2017targeted} sample a random image or a fixed
noise mask from uniform distribution as backdoor triggers, and construct backdoor images by mixing
up triggers with original images. This renders the
constructed poisoned images visually indistinguishable from benign ones, evading potential human inspection. While aforementioned backdoors choose manually
created features as backdoor triggers, triggers for semantic backdoors
\cite{bagdasaryan01} could be selected as any naturally occurring feature of the
physical world. The
adversary could also construct a special type of semantic backdoors, termed as
the edge-case backdoors \cite{Hongyi01}. This kind of backdoors adopts data that lives in the tail of
the input distribution as poisoned images. Such injected backdoors are less likely to conflict with
benign updates. This equips injected backdoors with stronger durability against the vanishing
backdoor effect \cite{chameleon,Zhengming01} and stealth against backdoor
detection mechanisms.

\subsection{Backdoor Defenses in FL}
Most existing backdoor defense mechanisms require the central server to process
received updates after clients finish local training. The server could try to
limit the influence of individual updates by either clipping them to a
predefined bound \cite{zitengsun01}, or adding random noises to the aggregated
model to interfere with potentially injected backdoors
\cite{McMahan2017LearningDP}. However, it is generally believed that solely
applying influence-reduction-based methods can only slow down the backdoor
injection rate, but disable from eliminating backdoors. Another line of
work tries to have the server identify backdoor updates among benign updates
by comparing model parameters. Identified anomaly updates are filtered
out from aggregation for the next global round. We then elaborate on introducing
several SOTA detect-then-filter-based methods in the following. Also, we select
these detection methods as baseline algorithms to demonstrate the effectiveness
of the proposed method in defending against long-term backdoor injections.

The server could equip the FL system with the byzantine-robust aggregation
protocol to filter out
backdoor updates. The baseline algorithm, Multi-Krum \cite{blanchard2017machine},  could ensure the
convergence of the distributed training with $n$ participants and at most $f$
malicious clients. For every received update, the server identifies
$n-f-1$ updates which are the closest in $\ell_2$ norm. The summation
of the computed $n-f-1$ distances is assigned to every update, and the one
with the smallest value is selected iteratively until $m$ updates are selected.
The server then aggregates all selected updates as the global model for
the next round.

A major line of detection-based methods tries to isolate backdoor updates from
benign updates by computing statistical metrics on received model
parameters. These methods generally build upon the assumption that
introducing backdoors makes poisoned models distinguishable from benign updates
in the parameter space. 
Deepsight
\cite{rieger2022deepsight} detects backdoor attacks by measuring differences in model updates using two metrics: Division Differences (DDifs) and Normalized Energy Updates (NEUPs). DDifs compare prediction scores between local and global models on random inputs, with deviations indicating poisoned models. NEUPs analyze parameter updates to infer label distributions. Local updates are classified as benign or poisoned based on NEUP thresholds and then clustered using NEUPs, DDifs, and cosine similarity. Clusters with a high ratio of benign models are accepted for aggregation. Foolsgold
\cite{foolsgold} focuses on securing FL in the sybil setting, where multiple
clients could be controlled by a single adversary. The defense mechanism is
built upon that benign updates are more diverse than poisoned ones,
as they share the same training objective. Different from directly ruling out
potential backdoors, Foolsgold assigns low aggregation weight to updates with
large cosine similarity with others to mitigate backdoor injections. FLAME
\cite{nguyen2022flame} first rules out suspicious updates through clustering
based on cosine similarity. All accepted updates are then clipped to the median
of the norm of all accepted models $m$. Finally, the server adds a sufficient
amount of Gaussian noise $\mathcal{N}(0,\sigma^2)$ to the aggregated model to
eliminate the injected backdoors, where
$\sigma=\frac{m}{\epsilon} \cdot \sqrt{2\ln\frac{1.25}{\delta}}$ for privacy budget
$\epsilon$ and $\delta$.

However, a recent work, Backdoorindicator, reveals the inherent weakness of methods that relies on
statistically examining received model parameters
\cite{backdoorindicator}. That is, these detection methods fail to
identify backdoor updates, which are fabricated statistically close to benign
ones. This work further combines proactive processing on the global model with post-training detection to enhance the detection success rate. This method
has the server first inject a pseudo-backdoor task, termed as the indicator
task, into the global model utilizing OOD data before broadcasting to local
clients. Once adversaries upload poisoned models, the indicator task will be
triggered and help the server rule out backdoor updates from aggregation.

Despite being the SOTA backdoor detection theme, BackdoorIndicator has
relatively poor performance when identifying backdoor updates in earlier
training stages. The undetected backdoor updates could still be incorporated
into the global model, rendering eliminating backdoors impossible. Considering
the difficulty of identifying every backdoor update for a long term, we propose
a novel backdoor defense mechanism, TrojanDam, which relies on robustifying the
FL global model leveraging OOD data before broadcasting to clients.
We then
elaborate on the details in the following.

%% file: motivation_method.tex
\section{TrojanDam}
In this section, we illustrate the rationale and detailed methodology of the
proposed method. We start by describing the threat model in terms of the goal
and capability of both the adversary and the defender. We proceed to reveal the
key intuition behind our method through considering why backdoors could be
planted into the FL global model. Through carefully designed experiments, we
find the decisive role of redundant neurons in successfully planting backdoors,
and further explore ways to robustify these neurons in a centralized machine
learning (ML) model leveraging OOD samples. Due to the privacy requirement of the FL
paradigm, we describe challenges when migrating the method to decentralized
settings and their corresponding solutions. Finally, we present the detailed
methodology of TrojanDam.

\subsection{Threat Model}

\noindent \textbf{Adversary's goal and capability.} The adversary tries to plant backdoors into the FL
global model by corrupting participating clients. Successfully injected
backdoors will make the FL model misclassify into an adversary-chosen label,
termed the target label, when encountering predefined backdoor triggers,
while leaving the main task accuracy uninfluenced. Once a local client is corrupted, the
adversary has access to its local dataset, and gains full control of the model
training and uploading process. The adversary could continuously participate
in the FL training process starting from any global round for a long range of
global rounds. We do not make constraints on types of injected backdoors. We assume a benign majority, where in each global round, the adversary could compromise up to 50\% of all participating clients.

\noindent \textbf{Defender's goal and capability.} The defender (also the FL
server) aims to learn a global model on the main task
$\{(x_m,y_m)|y_m\in\mathcal{Y}_m\}$, where $\mathcal{Y}_m$ is the label space of
the main task data, through iteratively broadcasting to a selected number of
clients, and aggregating received updates. Meanwhile, the defender will
try to protect the global model from backdoor injections. We
assume that the defender has no access to data that is in the same distribution as the raw data of participating clients, which adheres to the privacy
requirement of the FL framework. We further assume that the defender could
collect a number of OOD samples
$\{(x_o,y_{ro})|y_{ro}\in\mathcal{Y}_{ro}\}$ from the public dataset. The
collected OOD samples have distinct real label space with the main task data,
which means $\mathcal{Y}_m\cap\mathcal{Y}_{ro}=\emptyset$.

\subsection{Effective Backdoor Injection via Poisoning Redundant Neurons}
We first illustrate the key intuition behind our proposed defense through
exploring the reason why backdoors can be planted into the FL global model, and
how it could be utilized to launch more powerful attacks.

In centralized settings, it is generally believed that the over-parameterization
of deep neural networks renders adversaries the ability to plant backdoors into
the machine learning model \cite{qi2022revisiting, gu2020badnets,
chen2017targeted}. Specifically, previous studies \cite{denton2014exploiting,
NIPS2015_ae0eb3ee,gong2021defense,gong2022atteq} revealed that only a limited
number of parameters are closely related to the main task, while pruning the
rest parameters (considered to be redundant neurons) has minimal impact on the
accuracy of the main task. Adversaries could thus inject backdoors on these
redundant neurons by incorporating poisoned samples into the training
dataset. The described over-parameterization assumption also accounts for the sparse
nature of gradients in stochastic gradient descent (SGD), which further sheds
light on why backdoors can be injected into the FL global model, even
if adversaries control only a single participant. Relevant studies
\cite{stich2018sparsified, ivkin2019communication} empirically show that the
majority of the $\ell_2$ norm of the aggregated benign updates is concentrated
in a very small number of coordinates, while leaving most redundant parameters
largely unchanged. \emph{Backdoors planted on these redundant parameters are then
aggregated into the global model without interference from other benign updates.}

Further exploiting the redundant neurons can enable adversaries to launch more
powerful backdoor attacks. Neurotoxin \cite{Zhengming01} proposes to identify
parameters that are most frequently updated by benign updates, and exclude
these parameters when training poisoned models. This allows adversaries to
focus on planting backdoors on redundant neurons. Specifically, adversaries
first identify the top-$k\%$ coordinates of the benign gradients utilizing
their benign dataset. Adversaries could then update the malicious model by
computing gradients on the poisoned dataset. These gradients are then projected
onto the bottom-(100-$k$)$\%$ coordinates, leaving top-$k\%$ parameters which
are frequently updated by benign clients unchanged.

\begin{figure}[!h]
    \begin{center}
    \centerline{\includegraphics[width=\columnwidth]{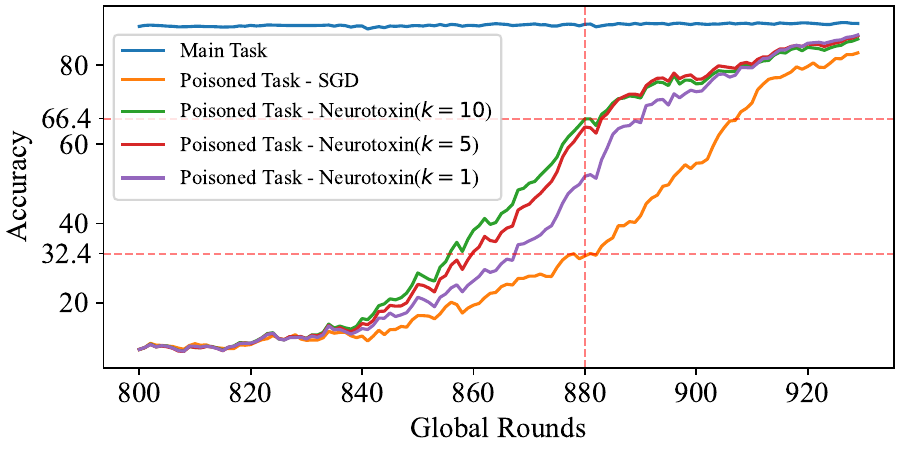}}
    \caption{Accuracies of FL global model on the main task and the poisoned
    tasks which are trained using SGD, and Neurotoxin with different percentages
    of excluded parameters ($k$). The adversary conducts single client attack in
    a continuous fashion staring from 800\textit{th} global round.}
    \label{motivation_neurotoxin}
    \end{center}
    \vspace{-0.3in}
\end{figure}

We empirically compare the backdoor performance between vanilla SGD and Neurotoxin with
different percentages of excluded parameters for backdoor injection.
We assume that the adversary, who tries to inject pixel-pattern
backdoors\cite{gu2020badnets}, successively controls only one out of all
selected clients in every global round starting from 800\textit{th} round. As shown in Figure \ref{motivation_neurotoxin}, the increase in the percentage
of excluded parameters achieves higher poisoned task accuracy on the FL global
model, when poisoning for the same number of global rounds. Specifically, injecting
backdoors utilizing Neurotoxin when excluding 10\% of coordinates achieves 66.4\%
poisoned task accuracy for the 880\textit{th} global round, which is over twice
of the poisoned task accuracy when only using SGD. This is because
deliberately excluding more frequently updated neurons causes backdoor information
to be planted more in redundant neurons. During FL aggregation, backdoor
information on redundant neurons hardly conflicts with benign updates, resulting
in high backdoor accuracy. The phenomenon further verifies the effectiveness of
injecting backdoors in FL by poisoning redundant neurons. \emph{It also indicates that
the effectiveness comes from a lack of conflict between benign and backdoor updates on
redundant neurons.}

Motivated by the special role of redundant neurons in planting backdoors into
the FL global model, \emph{we explore ways to defend against backdoor
attacks through robustifying redundant neurons in the FL global model. Specifically, we try to activate redundant neurons
by correlating them with additional information. Thus, the effects of backdoor updates on
redundant neurons can be canceled during aggregation with benign updates, whose redundant neurons remain activated.} In the following
section, we propose such a method that could activate redundant neurons, and
correlate them with out-of-distribution (OOD) data in centralized ML models.

\subsection{Activating Redundant Neurons Leveraging OOD Data}
The idea of activating redundant neurons utilizing OOD data is motivated by the
recently revealed property of backdoor tasks\cite{backdoorindicator}, which is
 \textbf{backdoor samples are essentially OOD samples concerning benign
samples from the target class.} Thus, the defender could first construct a
dataset with samples that are OOD concerning main task samples. The
defender could then simulate the behaviors of backdoor attackers, and inject a
sufficient number of OOD mappings into the model to activate redundant neurons.
We first elaborate on the proposed method in a centralized setting.

In the following, we consider a defender who aims to train a centralized ML
model, and tries to activate redundant neurons utilizing OOD data.
Specifically, we assume the defender is training ResNet18 on CIFAR10. In addition to main task samples $(x_m,y_m)$ from
CIFAR10, the defender has extra access to a limited number of OOD samples
$(x_o,y_o)$, which we assume are sampled from CIFAR100 wlog. To activate
redundant neurons, the defender then trains the model through computing updates
on the dataset $D_1=(\{(x_m^i,y_m^i)\}^N_{i=1},\{(x_o^j,y_o^j)\}^M_{j=1})$, which
is constructed by mixing up $N$ main task samples and $M$ OOD samples.
Corresponding labels for OOD samples are randomly assigned, and updated every a
fixed number of training iterations. This could help to ensure a continuously
sufficient number of OOD mappings for the ML model to learn over iterations. For comparison, we also consider two additional
scenarios where the defender constructs the training dataset using only main task
data, or only OOD data, which are $D_2=\{(x^i_m,y^i_m)\}^{N+M}_{i=1}$ and 
$D_3=\{(x^i_o,y^i_o)\}^{N+M}_{i=1}$ respectively. 

\begin{figure}[!h]
    \begin{center}
    \centerline{\includegraphics[width=\columnwidth]{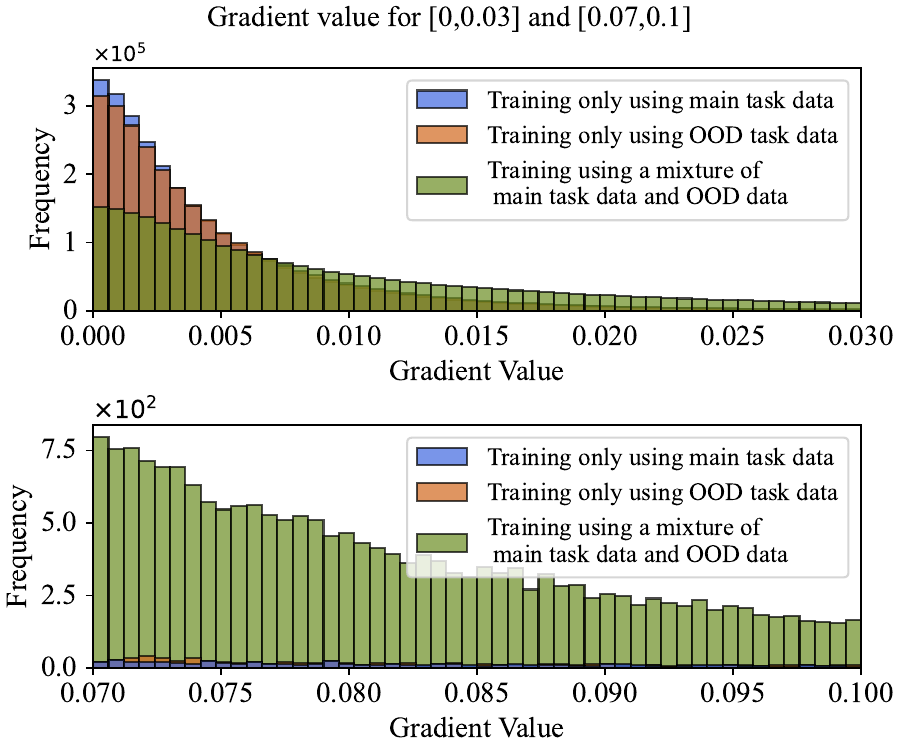}}
    \caption{The magnitude distribution of model gradients which are trained for
    a fixed number of iterations using \textbf{(GREEN)} a mixture of main task data and
    OOD data, \textbf{(BLUE)} only main task data, and \textbf{(ORANGE)} only OOD data.}
    \label{motivation_hist}
    \end{center}
    \vspace{-0.2in}
\end{figure}

Defenders in these scenarios proceed to train the ML model for the same number
of iterations with their datasets. When training finishes, model
gradients are reinitialized to zero, and recomputed using the same set of main task
data. We illustrate the magnitude distribution of gradients in Figure
\ref{motivation_hist}, which demonstrates that \emph{training by mixing up OOD
data and main task data could significantly activate redundant neurons in the
ML model}. Specifically, the magnitude of gradients training
using only main task data, or only OOD data is largely concentrated within 0.01.
These gradients contain few coordinates with value exceeding 0.02. While
training with a mixture of main task data and OOD data could reduce the number
of neurons with gradient lie in $[0,0.0005]$ by more than half, and
generate a considerable number of neurons with gradient larger than 0.03.
The gap is further exacerbated when considering the number of neurons with larger
gradients. The model trained using both main task data and OOD data
has around 800 neurons with 0.07 gradient, and even over 200 neurons
with 0.1 gradient. However, the number of such neurons within the models
trained using either only main task data or only OOD data are nearly 0. This
indicates that training with both OOD and main task data could help
to activate redundant neurons, while a lack of either OOD data or main task data
fails.

The effectiveness of activating redundant neurons utilizing OOD data in ML
settings motivates us to immigrate the method to FL settings, where the server
(as the defender) could continuously inject fresh OOD mappings into the global
model at the beginning of every global round. However, executing such a
mechanism in a decentralized manner faces several challenges due to the strict
privacy requirement of FL framework. We then elaborate on these challenges and
corresponding solutions in the next part.

\subsection{Challenges When Migrating to FL settings}
\label{subsection_challenges}
\textbf{1) The lack of main task samples on the FL server.}
Different from centralized settings where the defender has access to main task
samples, the server in FL has no access to the main task data in FL framework. The lack of main task samples disables the
server from imitating adversarial behaviors, making it hard to precisely
localize and activate redundant neurons by introducing OOD mappings. 
This could eventually lead to a
decrease in the backdoor suppression performance.

\textbf{2) Injecting a large number of OOD samples increases the
needed time for the robustification to take effect.} To effectively activate redundant neurons within the entire model, the
FL server needs to incorporate a sufficient number of OOD samples into training.
However, the increasing demand for OOD samples not only brings difficulty for the server to
collect, but also takes longer for the defense to take effect.

\begin{figure}[!h]
    \begin{center}
    \scalebox{0.95}{
    \centerline{\includegraphics[width=\columnwidth]{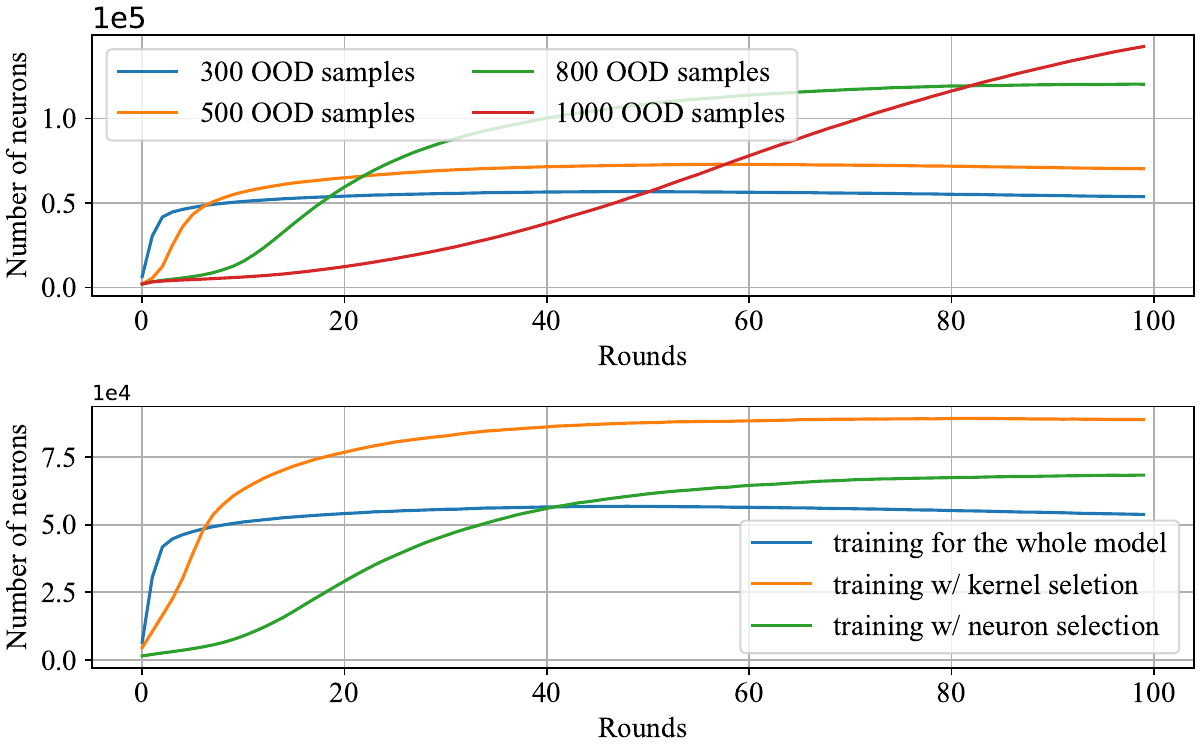}}
    }
    \caption{ The number of neurons with gradients larger than 0.05
    \textbf{(UPPER)} trained using different OOD dataset
    size, and \textbf{(LOWER)} trained using 300 flood
    samples, and different parameter selection methods.}
    \label{motivation_hist_challenge2}
    \end{center}
    \vspace{-0.2in}
\end{figure}

We demonstrate this challenge following the setting in Section 3.3,
where a centralized defender trains its ML model for certain rounds, using both main task and OOD data. For the constructed training dataset, we
fix the size of the main task dataset, and vary the number of OOD samples. We describe the
effectiveness of activating redundant neurons through the total number of
neurons (denoted as active neurons) with a gradient value larger than a certain
threshold (0.05 in this case). As shown in the upper part of Figure
\ref{motivation_hist_challenge2}, although more OOD samples could
effectively increase the number of active neurons, it takes more rounds to take
effect. Specifically, incorporating 1000 OOD samples into training increases
the number of active neurons to over $1.4\times10^5$ after 100 rounds, achieving
the strongest activating effect compared to a smaller OOD dataset size. However,
it merely gets $0.2\times10^5$ active neurons after 30 rounds, which is lower
than half of the number of active neurons trained using 300 OOD samples.
When migrating to FL settings, the server is unaware of the round when
adversaries start poisoning. Thus, it is desired to activate enough redundant
neurons as soon as possible to prepare the global model against backdoor
poisoning. 


In the following, we propose special designs to address the above challenges.
Besides the OOD dataset which is used to activate redundant neurons and termed
the \emph{flood dataset}, we suggest sampling an additional OOD dataset, termed
the \emph{shadow dataset} to substitute for the main task data. The global training dataset
is then constructed using a mixture of shadow and flood data. For the
tradeoff between large flood dataset size and short preparation time needed by the
FL server, 
we propose to only update the model on
\emph{a small set of key convolution kernels} which are the most vulnerable to backdoor
attacks and insensitive to benign training. Building upon the proposed
techniques, we develop a novel server-side backdoor defense method,
\emph{TrojanDam}\footnote{The proposed method compares to a dam that controls the
number of redundant neurons. The dam could release OOD mappings (like the flood)
to activate redundant neurons in favor of a robustified FL model.}, which could proactively suppress the backdoor injection through activating redundant neurons leveraging OOD data.
We then proceed to elaborate on the details of TrojanDam.

\subsection{Detailed Methodology}
To implement TrojanDam, the server starts with constructing the flood dataset
and the shadow dataset using OOD samples. At the beginning of every FL training
round, the server updates sample labels from both the flood
dataset and the shadow dataset. Specifically, labels in the flood dataset need
to be randomly assigned, and those in the shadow dataset need to be assigned
using the prediction of the current FL global model on corresponding shadow
samples. The server proceeds to kernel-wisely identify a small set of key
parameters using both the shadow and flood data. The training is then
conducted through computing updates, which are projected on the identified key
parameters, on a mixture of shadow data and OOD data. The server could then
broadcast the global model to all selected clients, and aggregate the received
updates for the next global round. The detailed algorithm is shown in Algorithm
\ref{algo_TrojanDam}. We next elaborate on each step in the following:

\begin{algorithm}[!h]
    \caption{TrojanDam}
    \label{algo_TrojanDam}
    \KwIn{OOD samples $(\{x_f^i\}_{i=1}^N,\{x_s^j\}_{j=1}^M)$, number of
    training iterations and learning rate in injecting OOD mappings: $E$,
    $\eta$, weight of the regularization term $\lambda$ and the ratio of the
    trainable parameters $\epsilon$. set of the selected local client at round
    $t$: $S_t$.} 
    \KwOut{Global model at global round $t+1$: $\boldsymbol{G}^{t+1}$} 

    \tcp{Server initializes at global round $t$}

    Randomly sample $n^i\sim U([-0.5,0.5]),\ y_f^i\sim U(\mathcal{Y}_m)$

    $D_f=\{(x_f^i+n^i,y_f^i)\}_{i=1}^N$

    $D_s=\{(x_s^i,y_s^i)\}_{i=1}^M,\ y_s^i=G^t(x_s^i)$

    $\boldsymbol{\mathcal{P}}=\text{IdentifyKeyKernels}(D_f, D_s, \epsilon)$

    The server saves estimated running mean and variance as $\boldsymbol{\mu}_M$
    and $\boldsymbol{\sigma}_M$.

    $\boldsymbol{\beta}^\prime\leftarrow \boldsymbol{G}^t$ 

    \For{$e=1,...,E$}{

        Compute stochastic gradient
        $\boldsymbol{g}_e=\nabla(\mathcal{L}_{task}(\boldsymbol{w}^\prime,D_f||D_s)+\lambda||\boldsymbol{w}^\prime-G^t||_2)$ 

        Project $\boldsymbol{g}_e$ on $\boldsymbol{\mathcal{P}}$ to get
        $\boldsymbol{g}_e^{\boldsymbol{\mathcal{P}}}$

        $\boldsymbol{\beta}^\prime=\boldsymbol{\beta}^\prime-\eta\boldsymbol{g}_e^{\boldsymbol{\mathcal{P}}}$
    }

    The server replace the BN statistics in $\boldsymbol{w}^\prime$ with $\boldsymbol{\mu}_M$
    and $\boldsymbol{\sigma}_M$.

    The server broadcasts $\boldsymbol{\beta}^\prime$

    \tcp{Clients perform local training}
    Clients initialize with $\boldsymbol{\beta}^\prime$

    Clients train their local models and update $\Delta
    \boldsymbol{\beta}_i=\boldsymbol{L}_i-\boldsymbol{\beta}^\prime$ to the server\\
    
    \tcp{Server aggregates}

    Clipping parameter $\gamma=\frac{1}{|S_t|}\sum_{i\in S_t}||\boldsymbol{\Delta\beta_i}||_2$

    $\boldsymbol{G}^{t+1} = \boldsymbol{\beta}^\prime+\frac{1}{|S_t|}\sum_{i\in
    S_t}\frac{\Delta\boldsymbol{\beta}_i}{max(1,||\Delta\boldsymbol{\beta}_i||_2/\gamma)}$
\end{algorithm}

\noindent \textbf{Constructing the flood dataset.} The sever first needs to
collect $N$ samples $\{x_f^i\}_{i=1}^N$, which are OOD concerning the main
task data, to construct the flood dataset. The server proceeds to additionally
sample $N$ noise masks $\{n^i\}_{i=1}^N$ with the same dimension of the collected
flood data. The flood dataset is then constructed by embedding flood samples with
corresponding noise masks. Their labels are uniformly drawn from
the main task label space. To keep the global model learning new
OOD mappings for continuously activating redundant neurons, the set of noise
masks, and labels should be resampled at the beginning of every FL
round. Specifically, we denote $\mathcal{Y}_m$ to be the main task label space,
the flood dataset $D_f$ is constructed and updated for every FL training round
such that
\begin{equation}
    D_f=\{(x_f^i+n^i,y_f^i)\}_{i=1}^N,\ y_f^i\sim U(\mathcal{Y}_m),\ n^i\sim U([-0.5,0.5])
\end{equation}

\begin{figure}[!h]
    \begin{center}
    \scalebox{0.8}{
    \centerline{\includegraphics[width=\columnwidth]{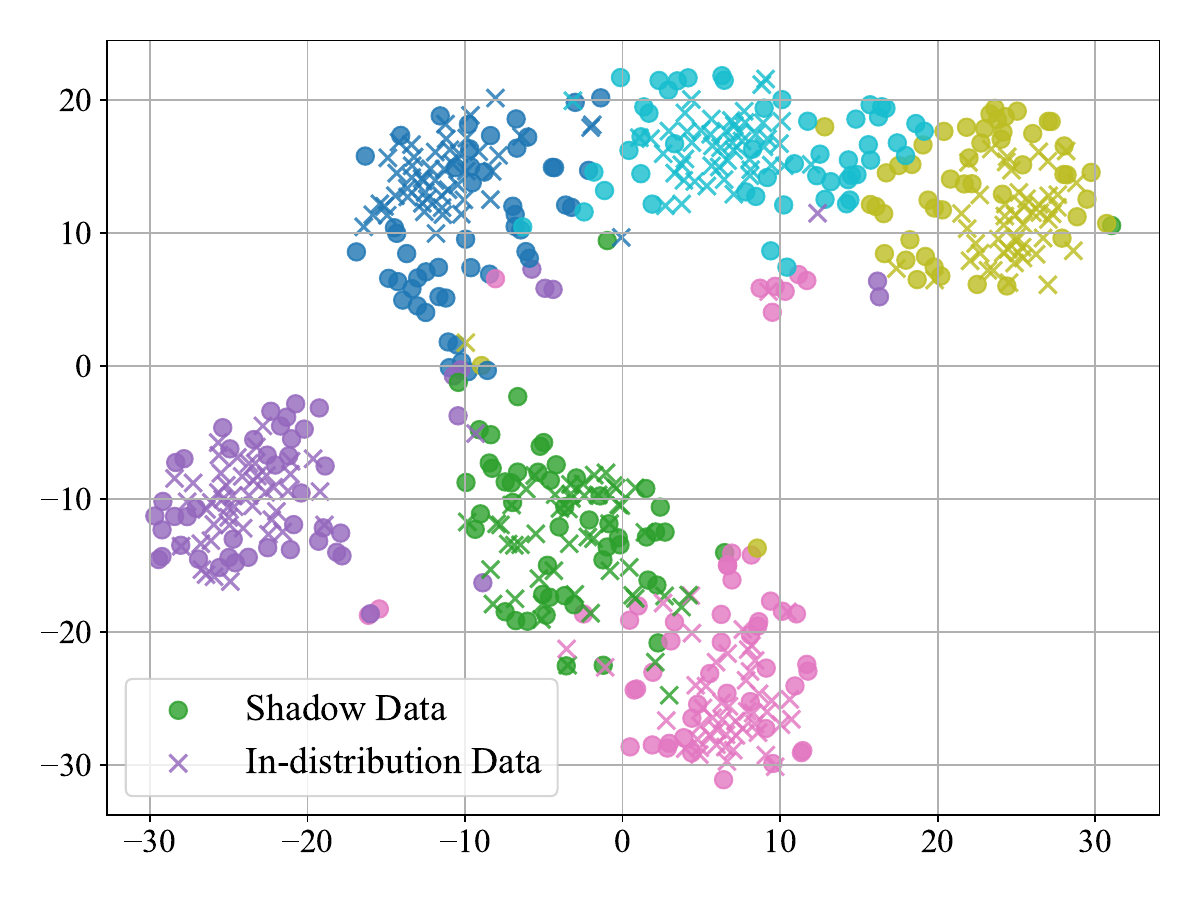}}
    }
    \caption{Feature space visualization of
    the shadow and in-distribution data with the same labels. Different colors
    represent different classes.}
    \label{shadow_data_justification}
    \end{center}
    \vspace{-0.1in}
\end{figure}

\noindent \textbf{Constructing the shadow dataset.} In addition to the flood
dataset, the server needs to sample another set of OOD samples
$\{x_s^i\}_{i=1}^M$ 
to construct the shadow dataset $D_s$. These shadow samples are then assigned with the
inference results 
from the current FL global model. This is because that ReLU neural networks tend to make overconfident
predictions on OOD samples. These models tend to assign OOD samples with
embeddings which are close to those of main task samples in the feature space
\cite{hein2019relu}. We further verify this property by visualizing the
shadow and in-distribution data in the feature space of a model which is
well-trained on CIFAR10. Empirically, we sample 300 images from 6 classes of
CIFAR10 with the number of images in each class equals to 50. For each
in-distribution sample, we construct its shadow counterpart by sampling an
image, which is predicted by the model as the label of its corresponding
in-distribution data, from CIFAR100. As it is shown in Figure
\ref{shadow_data_justification}, shadow data and in-distribution data, which
shares the same label, are clustered in the feature space. Thus, OOD samples
with model-assigned labels (termed as shadow data) could be considered as the
substitution for the main task data, and further incorporated into the training
to locate redundant neurons. 

We further require the label
distribution of the shadow dataset to be a uniform distribution of the main task
label space, which could avoid the potential distribution shift on the main task
accuracy. Specifically, we denote the FL global model as $\mathcal{F}$, the
shadow dataset $D_s$ is constructed at the beginning of every FL training round
in the following: 

\begin{equation}
    D_s=\{(x_s^i,y_s^i)\}_{i=1}^M,\ y_s^i=\mathcal{F}(x_s^i)
\end{equation}

\begin{algorithm}[!t]
    \caption{IdentifyKeyKernels}
    \label{algo_identifykeykernels}
    \KwIn{The flood dataset $D_f$, the shadow dataset $D_s$, and the ratio of the
    trainable parameters $\epsilon$.} 
    \KwOut{Selected trainable parameters $\boldsymbol{\mathcal{P}}$} 

    Compute gradients $\mathcal{G}_f=(\{(\boldsymbol{\mathcal{K}^i_f}, \gamma^i_f, \beta^i_f)\}_{i=1}^J,
    (\boldsymbol{w_f},\boldsymbol{b_f}))$ using $D_f||D_s$

    Compute gradients $\mathcal{G}_s=(\{(\boldsymbol{\mathcal{K}^i_s}, \gamma^i_s, \beta^i_s)\}_{i=1}^J,
    (\boldsymbol{w_s},\boldsymbol{b_s}))$ using $D_s$

    Compute average gradient different of convolution kernels
    $\{\{k^i=\text{Avg}(\boldsymbol{\mathcal{K}}_f^i-\boldsymbol{\mathcal{K}}_s^i)\}_{i=1}^J\}$

    $\boldsymbol{I}\leftarrow$Index list of kernels after ranking
    $\{k^i\}_{i=1}^J$ in a descending order

    $\boldsymbol{\mathcal{P}}=[\ ], ind=0$\tcc*[f]{Initialization}

    $N_m\leftarrow$ The total number of model parameters excluding the
    classifier

    \While{$\text{Len}(\boldsymbol{\mathcal{P}})/N_m<\epsilon$}{
        $\boldsymbol{p}\leftarrow$kernel with the $ind$-th largest $k$ ($k^{\boldsymbol{I}[ind]}$)

        $(\gamma_{\boldsymbol{p}},\beta_{\boldsymbol{p}})\leftarrow$ 
        corresponding BN scaling weight and bias of kernel $\boldsymbol{p}$

        $\boldsymbol{\mathcal{P}}\leftarrow \boldsymbol{\mathcal{P}}
        ||(\boldsymbol{p},\gamma_{\boldsymbol{p}},\beta_{\boldsymbol{p}})$

        $ind$ += 1
    }

\end{algorithm}

\noindent \textbf{Identifying key kernels.} We further propose to have the
server only update a small fraction of all model parameters to simultaneously
achieve a decent effect on activating redundant neurons and short preparation
time. The FL server could utilize the flood dataset and shadow dataset to identify
parameters, which are the most vulnerable to backdoor attacks and insensitive to
the main task. Also, we propose to select parameters based on the average
gradient of \textit{convolution kernels} instead of individual neurons. This is
because the convolution kernels serve as feature encoders in the model,
which possess stronger expressing ability and are therefore more vulnerable to
backdoor attacks compared to parameters in classifier and BN \cite{ioffe2015batch} layers. We further
justify this design in the following part.

Specifically, we assume that the server has constructed the flood dataset $D_f$,
and the shadow dataset $D_s$. The server proceeds to compute the model gradient
$\mathcal{G}_f=(\{(\boldsymbol{\mathcal{K}^i_f}, \gamma^i_f, \beta^i_f)\}_{i=1}^J,
(\boldsymbol{w_f},\boldsymbol{b_f}))$, using a mixture of the flood dataset and
the shadow dataset $D_f||D_s$. Here, we decompose the computed gradients into $J$
sets of convolution kernels, their corresponding BN scaling weights and biases $(\boldsymbol{\mathcal{K}_f}, \gamma_f, \beta_f)$. We further denote
the weight and bias of the subsequent classifier to be $(\boldsymbol{w_f},
\boldsymbol{b_f})$. The server also needs to compute the model gradient
$\mathcal{G}_s=(\{(\boldsymbol{\mathcal{K}^i_s}, \gamma^i_s, \beta^i_s)\}_{i=1}^J,
(\boldsymbol{w_s},\boldsymbol{b_s}))$ using only the shadow dataset $D_s$. 
The large difference between $\boldsymbol{\mathcal{K}_f}$ and $\boldsymbol{\mathcal{K}_s}$ denotes the kernel is more vulnerable to be injected with backdoors, and more irrelevant to the main task.
Thus, the set of trainable parameters $\boldsymbol{\mathcal{P}}$ could be constructed by
iteratively incorporating the kernel with the largest $k^i=\text{Avg}(\boldsymbol{\mathcal{K}}_f^i-\boldsymbol{\mathcal{K}}_s^i)$, and its
corresponding $\gamma^i$ and $\beta^i$ into $\boldsymbol{\mathcal{P}}$ until
the percentage of total trainable parameters reaches the limit $\epsilon$. Notably, we only select candidate
parameters from convolution kernels and their corresponding BN parameters, while
leaving classifier parameters $(\boldsymbol{w},\boldsymbol{b})$ unchanged. The
detailed algorithm for constructing the trainable parameter set
$\boldsymbol{\mathcal{P}}$ is shown in Algorithm 2.


\noindent \textbf{Design justification of updating key kernels.} We further
provide empirical results to demonstrate the effectiveness of updating key
kernels instead of individual neurons. Following the setting in Section 3.3, we
consider the defender updates its ML model on parameters identified using
Algorithm 2 with $\epsilon=0.15$. We consider another setting where the defender
directly identifies parameters that rank top-15\% in the difference between the
gradient obtained using a mixture of the flood dataset and the shadow dataset, and
that calculated using only the shadow dataset. 


In the lower part of Figure \ref{motivation_hist_challenge2}, updating
models with 300 flood samples on parameters selected on kernels achieves the
strongest activating effect across all evaluated settings. It activates over
$8\times10^4$ active neurons, which is about twice when
training the whole model or poisoning on selected neurons with the same flood dataset size. Updating models on key kernels could also shorten the
preparation time to quickly equip the model with the ability against potential
backdoor adversaries. It achieves a promising activating effect in 20 training
rounds, and surpasses all other settings after training for only 8 rounds.

\noindent \textbf{Injecting OOD mappings.} Having constructed the flood dataset
$D_o$, the shadow dataset $D_s$ and the trainable parameter set
$\boldsymbol{\mathcal{P}}$, the server could then begin to inject OOD mappings
to activate redundant neurons. At the beginning of every FL training round $t$,
the server first needs to update the flood dataset by refreshing the noise
mask set and assigning new random labels. Labels in the shadow dataset also
need to be updated by the prediction results of the current global model on
shadow samples. Then, the server saves the current BN statistics, and then
constructs the training dataset $D_t=D_s||D_f$ through mixing the flood dataset and
the shadow dataset. The global model is then trained through computing updates
on $D_t$ via optimizing the cross-entropy loss $\mathcal{L}_{\text{task}}$. To
further control the influence of injecting OOD mappings on the main task
accuracy, we regularize the model through punishing updates that deviate too
much from the original model. Specifically, let $w^\prime$ be the training
global model, and $G^t$ be the original global model. The server minimizes the
following loss with respect to $D_t$:
\begin{equation}
    \mathcal{L} = \mathcal{L}_\text{task}+\lambda||w^\prime-G^t||_2,
\end{equation}
where $\lambda$ denotes the weight of the regularization term. The computed
updates are then projected on the selected trainable parameters
$\boldsymbol{\mathcal{P}}$. The server then replaces the BN statistics with the
previously saved ones to avoid BN statistic drift
\cite{backdoorindicator}. The robustified global model is then broadcast
to all selected clients. After receiving updates from clients, the server clips
all updates to an adaptive bound, which is computed as the average of the norm of
all received updates. The initial global model for $t+1$ round is then generated
through aggregating all clipped updates.

%% file: exp.tex
\section{Experiments}

\begin{table*}[htbp]
    \centering
    \caption{BA (MA) of single client attack with different
    combinations of backdoor type and malicious training algorithm against all
    evaluated defense mechanisms. \textbf{Bold} values indicate
    the lowest metrics, while \underline{underlined} denotes the second
    lowest.}
    \scalebox{0.75}{
    \begin{tabular}{cc||cccccccccc}
    \toprule
    training alg. & bkdr. types & Nodefense & Deepsight & Foolsgold & FLAME & FreqFed & BayBFed & MESAS & FLTrust & Indicator & TrojanDam \\
    \midrule
    \multirow{4}[2]{*}{PGD} & blended & 78.73 (89.69) &
    \underline{41.7} (90.01) & 75.9 (88.42) & 89.02 (87.69) & 79.64 (88.52) &
    83.25 (90.03) & 77.65 (89.71) & 77.05 (90.36) &
        43.32 (90.46) & \textbf{9.79} (88.50) \\
        & semantic & 65.32 (90.43) & 69.61 (90.27) & 69.69 (89.89) &
        66.87 (88.30) & 69.14 (87.47) & 78.02 (88.26) & 79.67 (87.91) & 65.55
        (88.95) & \underline{52.13} (90.35) & \textbf{17.85} (88.03)
        \\
        & edge case & 70.39 (90.44) & 52.73 (90.16) & 64.28 (90.36) &
        81.92 (88.18) & 69.89 (88.88) & 69.73 (89.04) & 70.46 (88.19) & 60.99
        (89.35) & \underline{10.57} (89.76) & \textbf{10.38} (88.84)
        \\
        & TacT & 96.34 (88.89) & 90.18 (88.37) & 96.22 (88.01) &
        99.36 (86.21) & 91.89 (87.43) & 83.87 (88.11) & 92.67 (89.21) & 93.70
        (88.77) & \textbf{0.84} (89.88) & \underline{1.30} (87.21)
        \\
    \midrule
    \multirow{4}[2]{*}{Neurotoxin} & blended & 79.49 (91.04) & 42.44 (89.58) &
    75.42 (89.51) & 88.84 (88.84) & 72.34 (88.18) & 84.62 (89.14) & 80.38
    (90.02) & 76.18 (89.70) & \underline{46.60} (90.22) &
    \textbf{10.00} (87.92) \\
        & semantic & 64.42 (90.27) & 67.40 (90.34) & 68.37 (89.63) &
        67.40 (88.68) & 77.25 (86.93) & 80.41 (88.63) & 89.41 (87.57)  & 66.51
        (89.31) & \underline{51.84} (89.49) & \textbf{14.92} (89.42)
        \\
        & edge case & 67.27 (91.09) & 53.57 (90.11) & 58.73 (89.81) &
        82.96 (88.47) & 58.03 (89.10) & 70.17 (88.58) & 61.23 (85.47) & 56.83
        (89.47) & \underline{18.65} (90.19) & \textbf{12.27} (88.30)
        \\
        & TacT & 95.11 (88.33) & 88.68 (88.84) & 96.27 (88.15) &
        \textbf{0.25} (88.30) & 71.15 (87.12) & 69.83 (87.56) & 86.60 (88.20) &
        93.47 (88.42) & 1.54 (89.69) & \underline{1.10} (87.93)
        \\
    \midrule
    \multirow{4}[2]{*}{Chameleon} & blended & 78.91 (90.11) & 43.74 (89.91) &
    74.66 (89.78) & 88.99 (88.42) & 49.87 (89.22) & 81.45 (88.95) & 80.61
    (90.11) & 72.55 (89.34) & \underline{40.77} (90.05) &
    \textbf{22.60} (89.27) \\
        & semantic & 63.32 (90.36) & 61.34 (89.98) & 61.40 (90.44) &
        63.41 (88.67) & 81.36 (88.20) & 82.54 (89.54) & 83.27 (86.24) & 61.81
        (88.79) & \underline{44.34} (89.44) & \textbf{16.52} (89.59)
        \\
        & edge case & 51.66 (90.30) & 39.75 (89.90) & 41.47 (88.26) &
        73.66 (89.67) & 26.53 (89.17) & 58.76 (88.77) & 59.95 (87.82) & 49.40
        (88.94) & \textbf{12.89} (90.69) & \underline{20.67} (88.10)
        \\
        & TacT & 97.35 (89.45) & 78.04 (88.73) & 94.16 (90.02) &
        99.38 (88.39) & 76.51 (87.89) & 67.80 (87.57) & 94.50 (87.26) &
        93.40 (88.49) & \underline{1.85} (90.14) & \textbf{1.48} (89.03) \\
    \midrule
    \multirow{1}[2]{*}{CerP} & optimized & 79.09 (90.18) & 45.81 (88.96) &
    74.27 (88.33) & 90.99 (88.37) & 72.15 (89.40) & 79.26 (89.21) & 77.42
    (88.80) & 63.80 (88.15) & \underline{41.69} (89.42) &
    \textbf{30.76} (87.28) \\
    \midrule
    \multirow{1}[2]{*}{PFedBA} & optimized & 97.47 (87.16) & \underline{82.68} (88.78) &
    96.70 (87.12) & 98.99 (87.82) & 96.77 (84.88) & 96.22 (86.34) & 99.73
    (87.22) & 99.93 (87.04) & 96.57 (86.71) &
    \textbf{35.97} (86.62) \\
    \bottomrule
    \end{tabular}%
    }
    \vspace{-0.1in}
    \label{tab:single_client_attack}%
  \end{table*}%


\subsection{Experimental Setup}
\noindent \textbf{System settings.} We implement an FL system using FedAVG
\cite{mcmahan2017communication} with a single machine using a NVIDIA RTX A6000.
Experiments are conducted on three computer vision datasets: CIFAR10, CIFAR100
\cite{krizhevsky2009learning} and EMNIST \cite{cohen2017emnist} using three
model architectures: VGG16 \cite{simonyan2014very}, ResNet18 and ResNet34
\cite{he2016deep}. We assume totally 100 local clients participating in the FL
system, among which 10 clients are randomly selected to contribute to every
global round. The training dataset is randomly partitioned over clients in a
non-IID fashion using Dirichlet sampling \cite{hsu2019measuring}, with the
sampling parameter $\alpha$ set to 0.9 by default. We also vary $\alpha$ to
evaluate presented methods under more challenging non-IID settings. 
Codes are available at \url{https://github.com/ybdai7/TrojanDam-backdoor-defense}.

\noindent \textbf{Adversarial settings.} We evaluate the effectiveness of the
proposed method against a strong adversary capable of crafting backdoor updates
using a range of malicious training algorithms, including
PGD~\cite{zitengsun01}, Neurotoxin~\cite{Zhengming01}, and
Chameleon~\cite{chameleon}. The adversary can also inject various types of
backdoors, such as blended backdoors~\cite{chen2017targeted},
TaCT~\cite{tang2021demon}, semantic backdoors~\cite{bagdasaryan01}, and
edge-case backdoors~\cite{Hongyi01}. For TaCT backdoors, target samples are
drawn from class 8, and, without loss of generality, all backdoors are assigned
a target label of 3. In the case of semantic backdoors, we use the
car-with-vertically-striped-walls-in-the-background from CIFAR10.
Additionally, adversaries may deploy optimized trigger attacks using
CerP~\cite{lyu2023poisoning} and PFedBA~\cite{lyu2024lurking}. Beyond
single-client attacks, the adversary may also control multiple clients. In
multi-client scenarios, attackers may either corrupt several clients to jointly
inject blended backdoors via Neurotoxin or employ DBA~\cite{xie2019dba}, which
is tailored for collaborative poisoning. We assume the adversary initiates
backdoor injection early during training-a challenging scenario for existing
defenses~\cite{backdoorindicator}. Specifically, the attack begins at the 430th
global round and continues for an extended duration. For CIFAR10, the poisoning
lasts 600 rounds, except in the case of optimized trigger attacks, which
are evaluated over 100 rounds.

\noindent \textbf{Baseline defenses.} We evaluate the effectiveness of TrojanDam
against the aforementioned adversary by comparing it with several SOTA backdoor
defense methods, including Deepsight~\cite{rieger2022deepsight},
Foolsgold~\cite{foolsgold}, FLAME~\cite{nguyen2022flame},
FreqFed~\cite{fereidooni2023freqfed}, BayBFed~\cite{kumari2023baybfed},
MESAS~\cite{krauss2023mesas}, FLTrust~\cite{cao2020fltrust}, and
BackdoorIndicator~\cite{backdoorindicator}. For BackdoorIndicator, we adopt the
variant with adaptive norm clipping. In addition, we include a baseline setting
with no explicit defense mechanism beyond adaptive norm clipping, referred to as
\textit{Nodefense}. We assess the performance of all defense methods using two
metrics: the mean backdoor accuracy (BA) over the final 20 global rounds, and
the accuracy on the main task (MA), to account for any potential degradation in
utility. In all reported results, \textbf{bold} values indicate the best
performance (i.e., lowest metric), while \underline{underlined} values denote
the second-best.

\noindent \textbf{TrojanDam settings.} To deploy TrojanDam effectively, the FL
server must collect a set of OOD samples to construct the flood and shadow
datasets. For the CIFAR10 main task, we use samples from CIFAR100 as the source
for both flood and shadow datasets. For experiments on other datasets, we
instead use CIFAR10 to construct these datasets. By default, the flood dataset
contains 800 samples, and the shadow dataset contains 300 samples. The trainable
parameter ratio $\epsilon$ is set to 0.15, and the regularization weight is
fixed at 0.8. The server begins injecting OOD mappings into the global model
starting from the 400\textsuperscript{th} global round-30 rounds prior to the
onset of backdoor poisoning. Note that the 400\textsuperscript{th} round
corresponds to an early training phase, during which the main task accuracy is
only around 78\%. We did not choose to start poisoning at even earlier rounds as
the magnitudes of benign updates are still large, and the effect of backdoor
injection is rather weak. In addition, we provide empirical results to analyze
the impact of key hyperparameters on the defense performance.

\subsection{Results}

\begin{table*}[htbp]
    \centering
    \caption{
    BA (MA) of single client attack under different non-IID settings, and different
    poisoned learning rates (\textit{plr}s) against all evaluated defense mechanisms. The
    backdoor type, and malicious training algorithms are blended
    backdoors and Neurotoxin.}
    \scalebox{0.8}{
      \begin{tabular}{cc||cccccccccc}
      \toprule
      alpha & \textit{plr}. & Nodefense & Deepsight & Foolsgold & FLAME & FreqFed &
      BayBFed & MESAS & FLTrust & Indicator & TrojanDam \\
      \midrule
      \multirow{3}[2]{*}{0.9} & 0.01  & 69.31 (90.48) & 42.20 (90.77) &
      63.94 (90.77) & 83.16 (89.92) & 71.61 (90.31) & 73.77 (89.62) & 65.41 (87.49) & 64.83 (90.41) & \underline{36.67} (90.16) &
      \textbf{11.73} (88.95) \\
            & 0.025 & 79.49 (90.43) & 42.44 (90.95) & 75.42 (90.76) &
            88.84 (89.98) & 72.34 (90.30) & 84.62 (89.39) & 80.38 (88.10) & 76.18 (90.39) & \underline{46.60} (90.06) &
            \textbf{10.00} (88.99) \\
            & 0.04  & 85.06 (90.36) & 42.42 (91.09) & 82.05 (90.76) &
            \underline{12.26} (90.43) & 55.32 (90.43) & 88.26 (90.20) & 83.64 (88.75) & 82.32 (90.45) & 63.58 (90.09) &
            \textbf{11.63} (89.23) \\
      \midrule
      \multirow{3}[2]{*}{0.5} & 0.01  & 72.69 (89.57) & 45.87 (89.33) &
      63.07 (89.06) & 84.98 (88.61) & 75.03 (88.76) & 68.75 (89.16) & 74.16 (88.79) & 61.86 (89.06) & \underline{43.14} (90.11) &
      \textbf{14.45} (88.48) \\
            & 0.025 & 82.24 (89.46) & \underline{43.69} (88.18) &
            74.46 (89.81) & 90.24 (88.43) & 81.71 (88.71) & 77.82 (88.35) & 82.66 (88.81) & 74.79 (89.04) & 67.35 (90.36) &
            \textbf{17.12} (87.65) \\
            & 0.04  & 87.67 (89.35) & \underline{49.60} (89.41) &
            80.88 (89.80) & 92.76 (88.41) & 79.03 (88.66) & 86.85 (88.48) & 84.89 (87.66) & 81.56 (89.02) & 80.39 (90.35) &
            \textbf{23.45} (88.70) \\
      \midrule
      \multirow{3}[2]{*}{0.2} & 0.01  & 48.42 (87.85) & 37.11 (88.35) &
      45.30 (86.52) & 78.28 (84.58) & 64.19 (84.96) & 63.84 (84.50) & 67.54 (83.36) & 41.05 (85.58) & \underline{14.21} (86.20) &
      \textbf{8.09} (85.74) \\
            & 0.025 & 66.13 (88.09) & 39.61 (88.41) & 45.15 (86.48) &
            86.93 (84.89) & 69.51 (84.16) & 78.24 (85.52) & 80.41 (82.87) & 55.77 (85.63) & \underline{32.29} (86.65) &
            \textbf{13.76} (85.85) \\
            & 0.04  & 78.03 (88.18) & 40.65 (88.52) & 69.74 (86.81) &
            \textbf{13.5} (85.29) & 62.36 (84.39) & 85.70 (84.51) & 86.39 (83.05) & 66.75 (85.60) & 52.17 (87.19) &
            \underline{16.25} (86.30) \\
      \bottomrule
      \end{tabular}%
      }
      \vspace{-0.1in}
    \label{tab:alphas_plrs}%
  \end{table*}%

\noindent \textbf{Performance against single-client attacks.} We evaluate the
effectiveness of TrojanDam under various combinations of backdoor types and
malicious training algorithms, with results summarized in
Table~\ref{tab:single_client_attack}. As shown in the table, TrojanDam
successfully suppresses the BA to the lowest value in most settings against a
consistent backdoor injection for 600 global rounds. In the few scenarios where
TrojanDam does not achieve the lowest BA, its performance remains highly
competitive, with only marginal differences compared to the best-performing
method. For example, when defending against blended backdoors trained with PGD,
TrojanDam reduces the BA to 9.79\%, outperforming the second-best method,
Deepsight, by 31.9\%. In the case of TaCT backdoors, although BackdoorIndicator
achieves the lowest BA at 1.54\%, TrojanDam exhibits comparable performance,
limiting the BA to 1.30\%, with a negligible difference of less than 1\%.

\begin{figure}[!h]
    \begin{center}
      \scalebox{0.9}{
    \centerline{\includegraphics[width=\columnwidth]{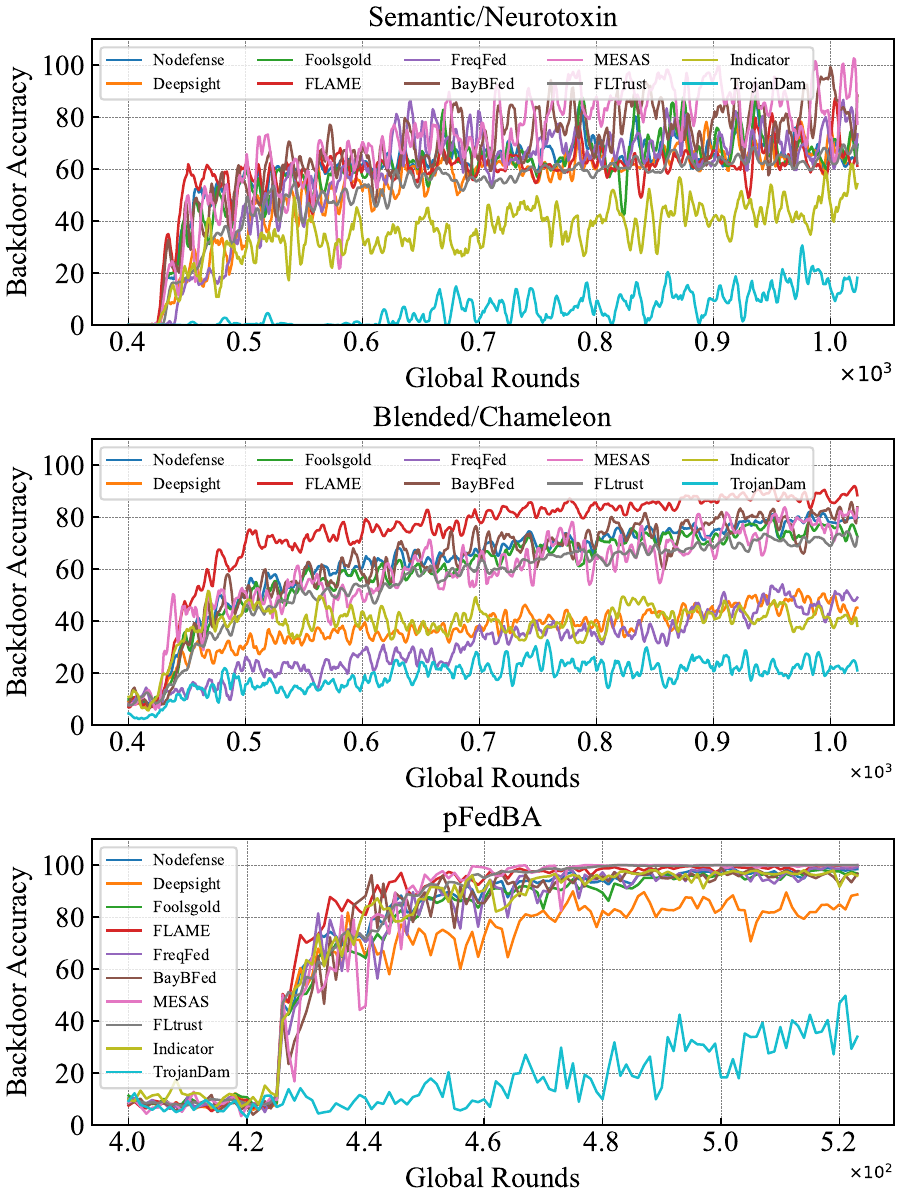}}
      }
    \caption{BA (MA) achieved by the injection of \textbf{(UPPER)}
    semantic backdoors trained using Neurotoxin, \textbf{(MIDDLE)} blended
    backdoors trained using Chameleon, and \textbf{(LOWER)} optimized trigger
    using PFedBA under various backdoor defense mechanisms.} 
    \label{fig:single_attack}
    \end{center}
    \vspace{-0.32in}
\end{figure}

For backdoor updates trained using more advanced malicious training algorithms,
like Neurotoxin and Chameleon, TrojanDam still effectively restricts the attack
success rate to nearly a random guess. TrojanDam achieves 14.92\% BA against the
injection of semantic backdoors trained using Neurotoxin, and 22.60\% BA against
Chameleon-trained blended backdoors. While the second-best defense mechanism
only gets corresponding backdoor accuracies of 51.84\% and 44.77\%, which are
over 20\% larger than those of TrojanDam. TrojanDam maintains its
effectiveness when facing optimized trigger attacks. For CerP,
TrojanDam achieves the lowest BA of 30.76\%, while BackdoorIndicator
achieves the second lowest BA of 41.69\% . Especially for the continuous attack of
PFedBA, other evaluated methods fail to defend for even 10 global rounds. As
shown in the lower part of Figure \ref{fig:single_attack}, the BA raises to
around 80\% at 440th global round for all other evaluated mechanisms. This is
due to the dual optimization process in PFedBA, generating both strong and
stealthy backdoor updates. However, TrojanDam could still restrict
the increase of BA to 35.97\% after 100 continuous backdoor injection,
demonstrating its strong backdoor mitigation performance. The
superior performance of TrojanDam against the injection of various backdoors
trained using different algorithms further demonstrates the adaptability of the
proposed method. 

TrojanDam maintains stable defense performance throughout the entire injecting
process. In Figure \ref{fig:single_attack}, TrojanDam consistently suppress the
BA to below 20\% for semantic backdoors trained using Neurotoxin, and blended
backdoors trained using Chameleon. The consistency of suppressing backdoor
injection over long iterations, together with the wide adaptability under
various adversarial settings strengthen the effectiveness of TrojanDam.

\noindent \textbf{Performance under different non-IID settings and different
poisoned learning rates.} Table \ref{tab:alphas_plrs} shows the performance of
all evaluated defense mechanisms under different non-IID settings, and different
\textit{plr}s. TrojanDam outperforms existing methods under different
non-IID settings, restricting the BA to around 15\%. Even for the
challenging data distribution setting where $\alpha=0.2$,
TrojanDam achieves 13.76\% BA when $\textit{plr}=0.025$, while
BackdoorIndicator achieves the second-best performance with 32.29\% BA.

As indicated in \cite{backdoorindicator}, adversaries may deliberately generate
backdoor updates using a small \textit{plr}, making them statistically
indistinguishable from benign updates and thereby evading statistical detection
mechanisms. Conversely, adversaries may also upload updates with large
deviations from benign ones by adopting a high \textit{plr}. These updates not
only remove the indicator tasks embedded by the server, but also significantly
amplify the impact of the backdoor on the global model.
Table~\ref{tab:alphas_plrs} presents the resilience of TrojanDam under varying
\textit{plr} settings. Although increasing \textit{plr} leads to a slight rise
in BA, TrojanDam remains robust, with the highest BA capped at 23.45\%, and most
cases restricted to around 15\%. For other defense mechanisms, we observe
that FLAME and FreqFed benefit from increased \textit{plr}, exhibiting improved
detection performance. This aligns with their design rationale, as both methods
rely on cosine similarity-either in the parameter space or frequency domain-to
identify anomalies. Backdoor updates trained with a large \textit{plr} tend to
deviate more from benign ones, making them more detectable. In contrast,
BackdoorIndicator becomes less effective under such conditions. Specifically,
when facing backdoor updates generated with 0.04 \textit{plr} and 0.5 $\alpha$,
the BA reaches 80.39\%. This degradation results from the gradual removal of the
server-planted indicator task, rendering BackdoorIndicator incapable of
detecting malicious behavior. Defense methods like Foolsgold, BayBFed, MESAS,
and FLTrust show similar trends across different \textit{plr} and $\alpha$
values. Their performance remains largely insensitive to $\alpha$, but they
gradually fail to defend against backdoor attacks as \textit{plr} increases.
This is due to the stronger influence of highly deviated backdoor updates on the
global model, which accelerates BA growth under these defenses. Deepsight
remains relatively stable, showing minimal sensitivity to both $\alpha$ and
\textit{plr}, with BA consistently around 40\% across all evaluated settings.

\begin{table}[htbp]
    \centering
    \caption{BA (MA) of multiple client attack against all
    evaluated methods. For Neurotoxin adversaries, the backdoor type
    is the blended backdoor.}
      \scalebox{0.8}
      {
      \begin{tabular}{c||cccc}
      \toprule
      training alg. & \multicolumn{3}{c}{Neurotoxin} & DBA \\
      \midrule
      bkdr. \% & 20    & 30    & 40    & 40 \\
      \midrule
      Nodefense & 87.05 (90.39) & 87.77 (90.16) & 89.03 (89.93) & 96.84 (88.96) \\
      Deepsight & 61.26 (90.95) & 67.58 (90.76) & 73.58 (90.62) & 90.37 (89.49) \\
      Foolsgold & \textbf{10.66} (90.99) & \underline{10.92} (90.82) & \underline{13.10} (90.79) & \textbf{10.38} (90.03) \\
      FLAME & 91.76 (89.31) & 93.25 (88.61) & 88.34 (77.33) & 99.07 (85.07)\\
      FreqFed & 50.90 (90.14) & 15.56 (90.26) & \textbf{12.91} (90.58) & 92.04 (89.97) \\
      BayBFed & 88.67 (89.69) & 90.22 (89.69) & 90.38 (89.40) & 93.60 (88.76) \\
      MESAS & 82.42 (89.27) & 88.81 (87.65) & 79.76 (87.38) & 89.20 (88.57) \\
      FLTrust & 84.31 (90.11) & 86.20 (89.92) & 86.32 (89.52) & 93.53 (88.95) \\
      Indicator & 53.91 (91.15) & 53.55 (90.78) & 50.94 (91.23) & 40.12 (90.43) \\
      TrojanDam & \underline{11.22} (88.82) & \textbf{10.29} (87.07) & 15.62 (88.06) & \underline{29.58} (88.35) \\
      \bottomrule
      \end{tabular}%
      }
    \label{tab:multiple_client_attack}%
    \vspace{-0.1in}
  \end{table}

\noindent \textbf{Performance against multiple client attack.} Table
\ref{tab:multiple_client_attack} demonstrates the defense performance of
different defense mechanisms against adversaries with the ability to control
multiple clients. For adversaries using Neurotoxin, Foolsgold and TrojanDam
consistently achieve both comparable and the lowest BA across all settings. As
it is shown in the table, these two mechanisms achieve 13.10\% and 15.62\% BA
respectively even if 40\% of all clients in every global round are compromised.
As Foolsgold is specifically designed for multiple client attacks, the comparable
performance between Foolsgold and TrojanDam further indicates the effectiveness
of TrojanDam against multiple client attacks. For adversaries controlling 40\% of
clients using DBA, Foolsgold achieves the lowest BA. TrojanDam restricts the BA
to 29.58\%, which is the lowest among all other methods. It is also
noteworthy that the performance of FreqFed is improved when the adversary
controls more clients to inject backdoors trained using Neurotoxin. FreqFed
achieves 50.90\% BA when 20\% of clients are compromised, and the BA drops to
12.91\% when 40\% of clients are compromised, which is the lowest among all
evaluated methods. However, it fails to defend against DBA adversaries, where
the BA increases to 92.04\% over long term backdoor injection. This is because
malicious updates trained using DBA have a strong influence on the global model.
The BA is quickly increased even if only few poisoned updates bypass the
detection. As shown in Figure \ref{fig:freqfed_dba} in the Appendix \ref{appendix},
the BA increases to 60\% after poisoning for around 300 global rounds, where the
percentage of the detected malicious updates by FreqFed maintains around 90\%.

\begin{table}[htbp]
    \centering
    \caption{BA (MA) achieved by performing single client attack with different model
    architectures and datasets against all evaluated methods. The poisoning lasts for 200 global rounds
    for EMNIST, 300 global rounds for VGG16 and 400 global rounds for ResNet34,
    CIFAR100.}
      \scalebox{0.8}
      {
      \begin{tabular}{c||cccc}
      \toprule
      arch./dataset & ResNet34 & VGG16 & CIFAR100 & EMNIST \\
      \midrule
      Nodefense & 68.18 (90.93) & 92.81 (90.11) & 65.20 (68.55) & 99.20 (99.71) \\
      Deepsight & 61.38 (90.76) & 85.56 (90.17) & \underline{27.76} (68.41) & 99.72 (99.69) \\
      Foolsgold & 70.95 (90.59) & 89.26 (89.92) & 63.78 (68.30) & 99.77 (99.70) \\
      FLAME & 69.58 (89.40) & 76.52 (88.47) & 63.78 (66.24) & 100.00 (99.69) \\
      FreqFed & \textbf{0.00} (90.12) & \underline{53.63} (89.96) & 60.92 (65.14) & 100.00 (99.65) \\
      BayBFed & 78.08 (89.74) & 73.16 (88.55) & 66.44 (66.63) & 100.00 (99.67) \\
      MESAS & 87.05 (90.15) & 94.51 (89.16) & 82.14 (66.38) & 100.00 (99.69) \\
      FLTrust & 59.96 (88.54) & 75.41 (90.41) & 82.14 (67.03) & 99.99 (99.67) \\
      Indicator & 26.85 (90.22) & 73.39 (90.65) & 52.30 (67.28) & \textbf{9.99} (99.70) \\
      TrojanDam & \underline{0.19} (89.22) & \textbf{0.48} (88.70) & \textbf{0.31} (65.55) & \underline{10.74} (99.48) \\
      \bottomrule
      \end{tabular}%
      }
    \label{tab:different_arch_dataset}%
    \vspace{-0.1in}
  \end{table}%


\noindent \textbf{Performance Across Model Architectures and Datasets.} We
evaluate TrojanDam under various model architectures and datasets. For
architecture comparisons, we adopt semantic backdoors and the Neurotoxin
training algorithm on CIFAR10. For dataset comparisons, we fix the backbone to
ResNet18, using blended backdoors with Neurotoxin on CIFAR100 and pixel-pattern
backdoors with Neurotoxin on EMNIST. Partial results are shown in
Table~\ref{tab:different_arch_dataset}, with full results available in
Appendix~\ref{appendix}. TrojanDam consistently achieves the best backdoor
suppression across architectures. Notably, it reduces the BA to 0.19\%,
comparable to FreqFed's 0\%. However, FreqFed's performance is unstable-its BA
increases to 53.63\% on VGG16, which is over 53\% higher than TrojanDam under
the same setting. This highlights TrojanDam's superior robustness and
adaptability to varying model architectures.

TrojanDam also exhibits strong adaptability across different datasets. On
CIFAR100, it achieves the lowest BA of 0.31\%, approaching the level of random
guessing. In contrast, the second-best method, Deepsight, reaches a BA of
27.76\%, while all other defenses result in BAs exceeding 50\%. On EMNIST, only
TrojanDam and BackdoorIndicator manage to reduce the BA to approximately 10\%,
whereas all remaining methods fail to suppress backdoor injection, with BAs
nearing 100\%.

\noindent \textbf{TrojanDam does not degrade the main task performance.} We
further present the MA when implementing TrojanDam against different backdoor
attacks. As shown in all listed tables, the MA of TrojanDam is comparable to
that of other defense methods across diverse settings. However, it still shows a
slight drop compared to scenarios without any defense. We proceed to analyze in
detail the impact of implementing TrojanDam on the MA in the following sections.

\subsection{Impact of Hyper-parameters}

\noindent \textbf{Influence of Flood and Shadow Dataset Sources.} We evaluate
the adaptability of TrojanDam under varying sources for both the flood and
shadow datasets. Specifically, we construct flood datasets by randomly sampling
from CIFAR100, EMNIST, and 300KRANDOM, each augmented with uniformly generated
noise. For shadow datasets, we restrict the sources to CIFAR100 and 300KRANDOM,
as EMNIST and random noise offer limited variability in the image space, making
it infeasible for the server to construct shadow datasets with distinguishable
labels.

\begin{table}[htbp]
    \centering
    \caption{BA (MA) achieved by 
    injecting blended backdoors using Chameleon against TrojanDam. The flood and
    shadow dataset are constructed using data from various sources.}
    \scalebox{0.9}{
    \begin{tabular}{c||cc}
    \toprule
    \diagbox{Flood}{Shadow} & CIFAR100 & 300KRANDOM \\
    \midrule
    CIFAR100 & 22.58$\pm$4.22 & 22.91$\pm$3.60 \\
    300KRANDOM & 28.59$\pm$3.02 & 22.90$\pm$3.15 \\
    EMNIST & 64.97$\pm$3.63 & 72.11$\pm$2.41 \\
    NOISE & 55.65$\pm$2.17 & 63.31$\pm$2.36 \\
    \bottomrule
    \end{tabular}
    }
    \label{tab:shadow_ood_source}
    \vspace{-0.1in}
  \end{table}

Table \ref{tab:shadow_ood_source} demonstrates that TrojanDam more effectively
suppresses BA when the flood dataset contains samples with
rich visual information. When using CIFAR100 as the flood source, TrojanDam
consistently limits BA to around 22\%, regardless of the shadow dataset.
Similarly, when flood samples are drawn from 300KRANDOM, BA remains low-22.90\%
with shadow samples from 300KRANDOM, and 28.59\% with CIFAR100. In contrast,
flood datasets constructed from EMNIST or random noise yield significantly
weaker defense: BA rises above 55\% in both cases. Specifically, using EMNIST as
flood data and CIFAR100 as shadow data results in a BA of 64\%. Random noise as
the flood dataset still offers moderate defense, achieving 55.65\% BA. These
results validate that richer feature representations in the flood dataset
improve TrojanDam’s effectiveness. EMNIST's binary-like pixel values (foreground
and background) lack the diversity required for robust suppression. Despite
performance degradation with random noise, TrojanDam still outperforms baselines
like Foolsgold (74.66\% BA) and FLAME (88.99\%) under identical settings.

\begin{table}[htbp]
  \centering
  \caption{Effective length and the MA of TrojanDam when
  using different sizes of the flood dataset. }
  \scalebox{0.95}{
    \begin{tabular}{c||cc}
    \toprule
    flood dataset size & Effective length & MA \\
    \midrule
    200   & 588   & 89.02$\pm$0.31 \\
    400   & 1874  & 87.62$\pm$0.35 \\
    600   & 1990  & 87.02$\pm$0.72 \\
    800   & 2104  & 86.67$\pm$0.70 \\
    \midrule
    NO ATTACK & - & 90.18$\pm$0.40\\
    \bottomrule
    \end{tabular}%
    }
    \vspace{-0.1in}
  \label{tab:ablation_size}%
\end{table}%

\noindent \textbf{Influence of the flood dataset size.} We empirically evaluate
how the size of the flood dataset affects backdoor defense performance, using
two metrics: 1) \textit{Effective length}, defined as the index of the first
global round where the global model's BA reaches 35\%, and 2) the mean and
variance of the MA over the final 20 global rounds. We also report the MA of
TrojanDam in a clean setting without adversaries for comparison. All experiments
are conducted under blended backdoor and Neurotoxin attack settings.

Table~\ref{tab:ablation_size} shows that increasing the size of the flood
dataset substantially enhances backdoor defense performance. For instance, with
only 200 samples, the BA remains below 35\% during the first
588 global rounds. Expanding the dataset to 400 samples extends the effective
length to 1874-approximately three times longer. Further enlarging the dataset
continues to improve the effective length, but the marginal gains diminish
compared to the initial increase from 200 to 400 samples. This is because a
larger flood dataset introduces more diverse feature representations, better
activating redundant neurons. However, this benefit comes at the cost of main
task performance. The MA with 200 samples reaches
89.02($\pm$0.31)\%, comparable to the no-attack setting. Increasing the dataset
to 400 and 800 samples results in an MA drop of 1.5\% and 3.5\%, respectively.
Once sufficient redundant neurons are activated, further expansion of the
dataset may interfere with the representation of the main task, leading to
accuracy degradation. Therefore, selecting an appropriate flood dataset size is
crucial to balance defense effectiveness and main task performance across
different scenarios.

\begin{table}[htbp]
  \centering
  \caption{Effective length and the MA of TrojanDam when
  using different key kernel ratios.}
  \scalebox{0.9}{
    \begin{tabular}{c||cc}
    \toprule
    Key kernel ratio & Effective length & MA \\
    \midrule
    0.10  & 1557  & 86.81$\pm$0.53 \\
    0.15  & 1910  & 87.24$\pm$0.74 \\
    0.20  & 1431  & 88.59$\pm$0.52 \\
    0.25  & 1369  & 89.13$\pm$0.30 \\
    0.15$^*$ & 2104  & 86.67$\pm$0.70 \\
    \midrule
    NO ATTACK & - & 90.18$\pm$0.40\\
    \bottomrule
    \end{tabular}%
    }
  \label{tab:ablation_ratio}%
  \vspace{-0.05in}
\end{table}%

\noindent \textbf{Influence of the Key Kernel Ratio.} We now investigate the
impact of the key kernel ratio on defense performance. The metrics used are the
effective length and MA, as described previously. The backdoor
types considered are blended backdoors and Neurotoxin. The default flood
dataset size is 400, with an additional setting of 800 flood samples denoted by
a $*$ to illustrate the relationship between the key kernel ratio and the flood
dataset size.

For the flood dataset with 400 samples, Table~\ref{tab:ablation_ratio} shows
that the defense performance improves as the key kernel ratio increases from
0.10 to 0.15, but degrades when further raised to 0.25. Incorporating more
kernels activates a greater number of redundant neurons, thereby enhancing
backdoor resistance. However, a higher key kernel ratio also demands more flood
samples to sufficiently activate these neurons. When the number of flood samples
remains fixed, increasing the ratio beyond 0.15 reduces the effective length
from 1910 to 1369 as shown in the table. This drop can be reversed by providing
more flood samples: for instance, increasing the sample count by 400 at a ratio
of 0.15 raises the effective length from 1910 to 2104. Moreover, a higher key
kernel ratio helps mitigate the negative impact of OOD mappings on main task
performance. With the same number of flood samples, the MA improves from 86.81\%
to 89.13\% as the ratio increases from 0.10 to 0.25, approaching the MA under
the no-attack condition.

\section{Resilience to Adaptive Attacks}
Powerful adversaries may adopt strategies to bypass the defense after knowing
how TrojanDam works. They could potentially 1) deliberately avoiding poisoning key
kernels, or 2) uploading updates with large norm to dominate the aggregation.

\noindent \textbf{Avoiding poisoning key kernels.} After knowing TrojanDam works
by robustifying key kernels, adversaries could simulate the procedure of
identifying key kernels and avoid poisoning them. We assume the advesary follows
Algorithm \ref{algo_identifykeykernels} using self-collected OOD data. When
poisoning local models, the adversary keep those identified key kernels
untouched. We test on three scenarioes where the server implements TrojanDam
with key kernel ratio equals to 0.15, 0.20 and 0.25, and the adversary keeps
corresponding percentage of key kernels untouched. The adversary injects blended
backdoors on CIFAR10 for 500 global rounds.

\begin{table}[htbp]
  \centering
  \caption{Performance of TrojanDam with different key kernel ratios against adaptive adversaries.}
  \scalebox{0.9}{
  \begin{tabular}{c||cc}
  \toprule
  Key kernel ratio & BA \\
  \midrule
  0.15 & 19.74$\pm$5.95 \\
  0.20 & 16.14$\pm$5.32 \\
  0.25 & 14.32$\pm$4.30 \\
  \bottomrule
  \end{tabular}%
  }
  \label{tab:adaptive_attacks_1}%
  \vspace{-0.2in}
\end{table}

As shown in Table \ref{tab:adaptive_attacks_1}, TrojanDam remains effective
against adaptive adversaries who avoid poisoning key kernels. For TrojanDam
implemented with key kernel ratio equals to 0.15, the final BA is 19.74\%. The
BA further drops to 14.32\% when 25\% of the key kernels are robustified. This
is because the identified key kernels are redundant neurons which interfere
little with other benign updates. Poisoning on other parameters rather than key
kernels faces frequent conflicts with the main task updates. Thus, this results
in the difficulty of injecting backdoors into the global model.

\noindent \textbf{Uploading updates with large norm.} The influence of maliciosu
updates with large norm could be effectively mitigated by the norm-clipping
component of TrojanDam. We further demonstrate the necessity of incorporating
norm clipping (NCD) into the whole defense mechanism. Here, we assume the
adversary try to continuously inject blended backdoors using Neurotoxin for 100
global rounds. The \textit{plr} is set to 0.1, which allows the adversary to
upload dominating malicious updates. The main task is CIFAR10, and the NCD takes
the median of the norm of all received updates as the threshold.
\begin{table}[htbp]
  \centering
  \caption{Performance of TrojanDam w./wo. NCD and only applying NCD against adaptive attack.}
  \scalebox{0.9}{
  \begin{tabular}{c||cc}
  \toprule
  method & BA \\
  \midrule
  only NCD  & 69.31$\pm$2.35 \\
  TrojanDam wo. NCD  & 52.47$\pm$4.35 \\
  TrojanDam w. NCD  & 18.9$\pm$3.59 \\
  \bottomrule
  \end{tabular}%
  }
  \label{tab:adaptive_attacks}%
\end{table}

As shown in table \ref{tab:adaptive_attacks}, the performance of TrojanDam is
significantly impaired without the incorporation of NCD. It only achieves
52.47\% BA, which is only around 17\% lower than that of only applying NCD.
However, combining TrojanDam with NCD could still constrict the BA to 18.9\%,
demonstrating its resilience to adaptive attacks.

%% file: discussion.tex




%% file: conclusion.tex
\section{Conclusion}

We propose a novel detection-free backdoor defense mechanism, TrojanDam, which
enhances the robustness of FL against continuous, long-term backdoor injections.
Unlike existing methods, TrojanDam eliminates the need to identify or filter
malicious client updates after aggregation. The core insight behind TrojanDam is
that successful FL backdoor attacks often exploit redundant neurons in the
global model. To counter this, TrojanDam proactively fortifies these neurons by
continually injecting OOD samples with randomly generated noise masks and
synthetic labels during server-side training. Extensive experiments across
diverse adversarial and systematic scenarios demonstrate the effectiveness,
adaptability, and practicality of TrojanDam, highlighting its potential for
real-world deployment.